\documentclass[pre,twocolumn,showpacs,superscriptaddress,preprintnumbers,floatfix]{revtex4-1}

\usepackage{etex}
\usepackage{ifpdf}
\usepackage{hyperref}
\usepackage{dcolumn}
\usepackage{url}
\usepackage{amsmath}
\usepackage{amscd}
\usepackage{amsfonts}
\usepackage{amssymb}
\usepackage{amsthm}
\usepackage{xfrac}
\usepackage{braket}
\usepackage{bm}   
\usepackage{bbm}
\usepackage{verbatim}
\usepackage{stmaryrd}
\usepackage{xcolor}
\usepackage{setspace}
\usepackage{tikz}
\usetikzlibrary{arrows}
\usetikzlibrary{automata}
\usetikzlibrary{positioning}
\usetikzlibrary{plotmarks}
\usepackage{pgfplots}
\pgfplotsset{compat=newest}

\usepackage{floatrow}
\tikzstyle{vaucanson}=[
  node distance=2.5cm,
  bend angle=15,
  auto,
  every loop/.style={},
  every edge/.style={->,draw=black,line width=1.2,>=latex,shorten <=1pt,shorten >=1pt},
  every state/.style={draw=black,line width=2,font=\large},
  loop right/.style={right,out=22,in=-22,loop},
  loop above/.style={above,out=112,in=68,loop},
  loop left/.style={left,out=202,in=158,loop},
  loop below/.style={below,out=292,in=248,loop},
  loop above right/.style={right,out=67,in=23,loop},
  loop above left/.style={left,out=157,in=113,loop},
  loop below left/.style={left,out=247,in=203,loop},
  loop below right/.style={right,out=337,in=293,loop},
]

\theoremstyle{plain}    \newtheorem{Lem}{Lemma}
\theoremstyle{plain}    
\theoremstyle{plain}    \newtheorem*{ProLem}{Proof}
\theoremstyle{plain}    \newtheorem{Cor}{Corollary}
\theoremstyle{plain}    
\theoremstyle{plain}    \newtheorem*{ProCor}{Proof}
\theoremstyle{plain}    \newtheorem{The}{Theorem}
\theoremstyle{plain}    
\theoremstyle{plain}    \newtheorem*{ProThe}{Proof}
\theoremstyle{plain}    
\theoremstyle{plain}    
\theoremstyle{plain}    
\theoremstyle{plain}    
\theoremstyle{plain}    \newtheorem*{Rem}{Remark}
\theoremstyle{plain}    \newtheorem{Def}{Definition}
\theoremstyle{plain}    \newtheorem*{Not}{Notation}
\theoremstyle{plain}

\newcommand{\eM}     {\mbox{$\epsilon$-machine}}
\newcommand{\eMs}    {\mbox{$\epsilon$-machines}}

\newcommand{\Process}{\mathcal{P}}

\newcommand{\MeasAlphabet}  {\mathcal{A}}
\newcommand{\MeasSymbol}   { {X} }
\newcommand{\meassymbol}   { {x} }
\newcommand{\MeasSymbols}[2]{ \MeasSymbol_{#1:#2} }
\newcommand{\meassymbols}[2] { \meassymbol_{#1:#2} }

\newcommand{\Past} { \MeasSymbols{}{0} }
\newcommand{\past} { \meassymbols{}{0} }

\newcommand{\Future} { \MeasSymbols{0}{} }

\newcommand{\CausalState}   { \mathcal{S} }

\newcommand{\causalstate}   { \sigma }
\newcommand{\CausalStateSet}    { \boldsymbol{\CausalState} }
\newcommand{\AlternateState}    { \mathcal{R} }

\newcommand{\AlternateStateSet} { \boldsymbol{\AlternateState} }

\newcommand{\Prob}      {\Pr} 

\newcommand{\Cmu}       {C_\mu}
\newcommand{\hmu}       {h_\mu}
\newcommand{\EE}        {{\bf E}}

\newcommand{\PC}        {\chi}

\newcommand{\CI}        {\Xi}

\newcommand{\ProcessAlphabet}   {\MeasAlphabet}

\newcommand{\forward}{+}
\newcommand{\reverse}{-}
\newcommand{\forwardreverse}{\pm} 

\newcommand{\FutureCausalState} { {\CausalState}^{\forward} }

\newcommand{\PastCausalState}   { {\CausalState}^{\reverse} }

\newcommand{\FutureCmu} { C_\mu^{\forward} }
\newcommand{\PastCmu}   { C_\mu^{\reverse} }

\newcommand{\lastindex}[2]{
  \edef\tempa{0}
  \edef\tempb{#2}
  \ifx\tempa\tempb
    \edef\tempc{#1}
  \else
    \edef\tempa{0}
    \edef\tempb{#1}
    \ifx\tempa\tempb
      \edef\tempc{#2}
    \else
      \edef\tempc{#1+#2}
    \fi
  \fi
  \tempc
}

\newcommand{\rhomu}{\rho_\mu}
\newcommand{\rmu}{r_\mu}
\newcommand{\bmu}{b_\mu}
\newcommand{\qmu}{q_\mu}

\newcommand{\sigmamu}{\sigma_\mu}

\newcommand{\CSjoint}[1][,]{
   \edef\tempa{:}
   \edef\tempb{#1}
   \ifx\tempa\tempb
      \ensuremath{\FutureCausalState\!#1\PastCausalState}
   \else
      \ensuremath{\FutureCausalState#1\PastCausalState}
   \fi
}

\newif\ifpm
\edef\tempa{\forwardreverse}
\edef\tempb{\pm}
\ifx\tempa\tempb
   \pmfalse
\else
   \pmtrue
\fi

\newcommand{\cs} {\causalstate}
\newcommand{\Measure}{ \mu }
\newcommand{\MIET}{ \mu + 1 }
\newcommand{\MIETI}{ \frac{1}{\mu + 1} }

\colorlet {R_color}    {blue}
\colorlet {k_color}    {black!30!green}

\def\clap#1{\hbox to 0pt{\hss#1\hss}}

\begin{document}

\title{Informational and Causal Architecture of\\
Discrete-Time Renewal Processes}

\author{Sarah Marzen}
\email{smarzen@berkeley.edu}
\affiliation{Redwood Center for Theoretical Neuroscience and
Department of Physics, University of California at Berkeley,
Berkeley, CA 94720-5800}

\author{James P. Crutchfield}
\email{chaos@ucdavis.edu}
\affiliation{Complexity Sciences Center and Department of Physics, University of
  California at Davis, One Shields Avenue, Davis, CA 95616}

\date{\today}
\bibliographystyle{unsrt}

\begin{abstract}
Renewal processes are broadly used to model stochastic behavior consisting of
isolated events separated by periods of quiescence, whose durations are
specified by a given probability law. Here, we identify the minimal sufficient
statistic for their prediction (the set of causal states), calculate the
historical memory capacity required to store those states (statistical
complexity), delineate what information is predictable (excess entropy), and
decompose the entropy of a single measurement into that shared with the past,
future, or both. The causal state equivalence relation defines a new subclass
of renewal processes with a finite number of causal states despite having an
unbounded interevent count distribution.
We use these formulae to analyze the output of the parametrized Simple
Nonunifilar Source, generated by a simple two-state hidden Markov
model, but with an infinite-state \eM\
presentation. All in all, the results lay the groundwork for analyzing
processes with infinite statistical complexity and infinite excess entropy.

\vspace{0.2in}

\end{abstract}

\preprint{Santa Fe Institute Working Paper 14-08-032}
\preprint{arxiv.org:1408.6876 [cond-mat.stat-mech]}

\maketitle


\setstretch{1.1}

\newcommand{\Abet}{\ProcessAlphabet}
\newcommand{\MS}{\MeasSymbol}
\newcommand{\ms}{\meassymbol}
\newcommand{\SSet}{\CausalStateSet}
\newcommand{\St}{\CausalState}
\newcommand{\st}{\causalstate}
\newcommand{\MxSt}{\AlternateState}
\newcommand{\MxSSet}{\AlternateStateSet}
\newcommand{\mxst}{\mu}
\newcommand{\mxstt}[1]{\mu_{#1}}
\newcommand{\StartMS}{\bra{\delta_\pi}}
\newcommand{\Ipred}{\EE}
\newcommand{\ISI} { \xi }

\newenvironment{entry}
  {\begin{list}{--}{
      \setlength{\topsep}{0pt}
      \setlength{\itemsep}{0pt}
      \setlength{\parsep}{0pt}
      \setlength{\labelwidth}{5pt}
      \setlength{\itemindent}{0pt}}}
   {\end{list}}


\section{Introduction}

Stationary renewal processes are widely used, analytically tractable, compact
models of an important class of point processes
\cite{Smit58a,Gerstner,Beichelt,Barbu}. Realizations consist of sequences of
events---e.g., neuronal spikes or earthquakes---separated by epochs of
quiescence, the lengths of which are drawn independently from the same
interevent distribution. Renewal processes on their own have a long history
and, due to their offering a parsimonious mechanism, often are implicated in
highly complex behavior \cite{Lowe93a,Lowe93b,Caki06a,Akim10a,Mont13a,Bolo13a}.
Additionally, understanding more complicated processes
\cite{Bian07a,Li08a,Kell12a,Onag14a} requires fully analyzing renewal processes
and their generalizations.

As done here and elsewhere \cite{Vale14a}, analyzing them in-depth from a
structural information viewpoint yields new statistical signatures of apparent
high complexity---long-range statistical dependence, memory, and internal
structure. To that end, we derive the causal-state minimal sufficient
statistics---the \eM---for renewal processes and then derive new formulae for
their various information measures in terms the interevent count distribution.
The result is a thorough-going analysis of their \emph{information
architecture}---a shorthand referring to a collection of measures that together
quantify key process properties: predictability, difficulty of prediction,
inherent randomness, memory, and Markovity, and the like. The measures include:
\begin{entry}
\item the \emph{statistical complexity} $\Cmu$, which quantifies the historical
	memory that must be stored in order to predict a process's future;
\item the \emph{entropy rate} $\hmu$, which quantifies a process' inherent
	randomness as the uncertainty in the next observation even given that we can
	predict as well as possible;
\item the \emph{excess entropy} $\EE$, which quantifies how much of a process's
	future is predictable in terms of the mutual information between its past
	and future;
\item the \emph{bound information} $\bmu$, which identifies the portion of the
	inherent randomness ($\hmu$) that affects a process's future in terms of the
	information in the next observation shared with the future, above
	and beyond that of the entire past; and
\item the \emph{elusive information} $\sigmamu$, which quantifies a process's
	deviation from Markovity as the mutual information between the past and
	future conditioned on the present.
\end{entry}
Analyzing a process in this way gives a more detailed understanding of its
structure and stochasticity. Beyond this, these information measures are key to
finding limits to a process's optimal lossy predictive features \cite{Shal99a,
Stil07a, Stil07b, Marz14a}, designing action policies for intelligent
autonomous agents \cite{Mart13a}, and quantifying whether or not a given
process has one or another kind of infinite memory
\cite{Crut01a,Debo12a,Trav11b}.

\begin{figure}[htp]
\includegraphics[width=\columnwidth]{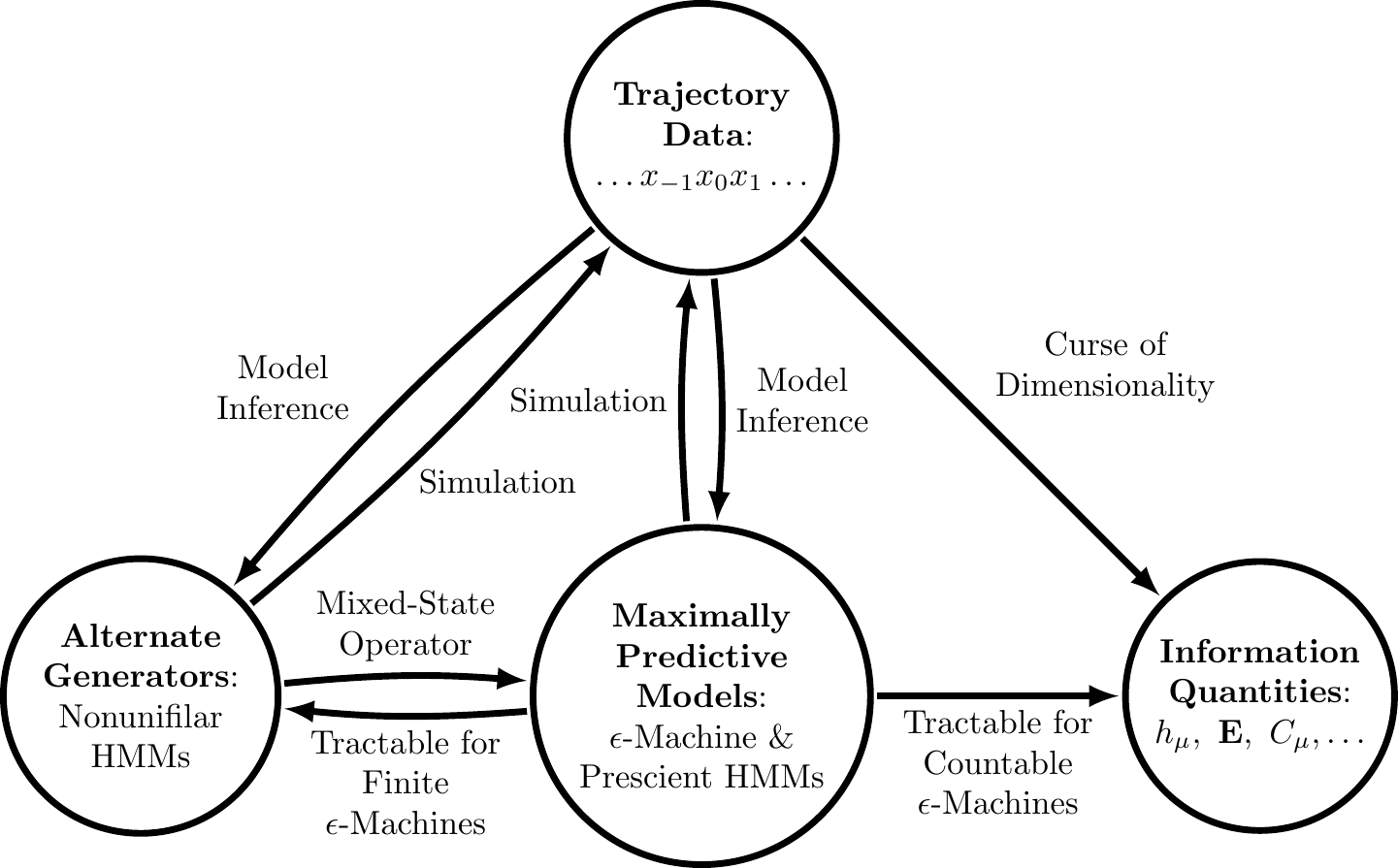}
\caption{\textbf{The role of maximally predictive (prescient) models:}
  Estimating
  information measures directly from trajectory data encounters a curse of
  dimensionality or, in other words, severe undersampling. Instead, one can
  calculate information measures in closed-form from (inferred) maximally
  predictive models \cite{Crut13a}. Alternate generative models that are
  \textit{not} maximally predictive cannot be used directly, as Blackwell
  showed in the 1950s \cite{Blac57b}.
  }
\label{fig:setup}
\end{figure}

While it is certainly possible to numerically estimate information measures
directly from trajectory data, statistical methods generally encounter a
\emph{curse of dimensionality} when a renewal process has long-range temporal
correlations since the number of typical trajectories grows exponentially (at
entropy rate $\hmu$). Alternatively, we gain substantial advantages by first
building a maximally predictive model of a process (e.g., using Bayesian
inference \cite{Stre13a}) and then using that model to calculate information
measures (e.g., using recently available closed-form expressions when the model
is finite \cite{Crut13a}). Mathematicians have
known for over a half century \cite{Blac57b} that alternative models that
are not maximally predictive are inadequate for such calculations. Thus,
maximally predictive models are critical. Figure~\ref{fig:setup} depicts the
overall procedure just outlined, highlighting their important role. Here,
extending the benefits of this procedure, we determine formulae
for the information measures mentioned above and the appropriate model structures for a class
of processes that require countably infinite models---the ubiquitous renewal
processes.

Our development requires familiarity with computational mechanics
\cite{Crut12a}. Those disinterested in its methods, but who wish to
use the results, can skip to Figs.~\ref{fig:Generic_eM1}-\ref{fig:Generic_eM4}
and Table \ref{tab:MeasuresFormulae}. A
pedagogical example is provided in Sec. \ref{sec:SNS}. Two sequels will use the
results to examine the limit of infinitesimal time resolution for information
in neural spike trains \cite{Marz14e} and the conditions under which renewal
processes have infinite excess entropy \cite{Marz14d}.

The development is organized as follows.
Section \ref{sec:Background} provides
a quick introduction to computational mechanics and prediction-related information measures of
stationary time series. Section \ref{sec:CS_DTSRP} identifies the causal states
(in both forward and reverse time), the statistical complexity, and the \eM\ of
discrete-time stationary renewal processes. Section \ref{sec:EM} calculates the
information architecture and predictable information of a discrete-time
stationary renewal process.
Section \ref{sec:SNS} calculates these
information-theoretic measures explicitly for the parametrized Simple
Nonunifilar Source, a simple two-state nonunifilar Hidden Markov Model with a
countable infinity of causal states.
Finally, Sec. \ref{sec:Conclusions} summarizes the results and lessons, giving
a view to future directions and mathematical and empirical challenges.


\section{Background}
\label{sec:Background}

We first describe renewal processes, then introduce a small piece of
information theory, review the definition of process structure, and finally
recall several information-theoretic measures designed to capture organization
in structured processes.

\subsection{Renewal Processes}
\label{sec:RenewalProcesses}

We are interested in a system's immanent, possibly emergent, properties. To
this end we focus on behaviors and not, for example, particular equations of
motion or particular forms of stochastic differential or difference equation.
The latter are important in applications because they are generators of
behavior, as we will see in a later section. As Fig.
\ref{fig:setup} explains, for a given process, some of its generators
facilitate calculating key properties. Others lead to complicated calculations
and others still cannot be used at all.

As a result, our main object of study is a \emph{process} $\Process$: the list
of all of a system's behaviors or realizations $\{ \ldots, \ms_{-2}, \ms_{-1},
\ms_{0}, \ms_{1}, \ldots \}$ as specified by their measure $\Measure(\ldots,
\MS_{-2}, \MS_{-1}, \MS_{0}, \MS_{1}, \ldots)$. We denote a contiguous chain of
random variables as $\MS_{0:L} = \MS_0 \MS_1 \cdots \MS_{L-1}$. Left indices
are inclusive; right, exclusive. We suppress indices that are infinite.  In
this setting, the \emph{present} $\MS_0$ is the random variable measured at $t
= 0$, the \emph{past} is the chain $\MS_{:0} = \ldots \MS_{-2} \MS_{-1}$
leading up the present, and the \emph{future} is the chain following the
present $\MS_{1:} = \MS_1 \MS_2 \cdots$. The joint probabilities
$\Prob(\MS_{0:N})$ of sequences are determined by the measure of the
corresponding cylinder sets: $\Prob(\MS_{0:N} = \ms_0 \ms_1 \ldots \ms_{N-1}) =
\Measure(\ldots, \ms_0, \ms_1, \ldots, \ms_{N-1}, \ldots)$. Finally, we assume
a process is ergodic and stationary---$\Prob(\MS_{0:L}) = \Prob(\MS_{t:L+t})$
for all $t \in \mathbb{Z}$---and the observation values $\ms_t$ range over a
finite alphabet: $\ms \in \Abet$. In short, we work with \emph{hidden
Markov processes} \cite{Ephr02a}.

Discrete-time stationary \emph{renewal processes} here have binary observation
alphabets $\Abet = \{0,1\}$. Observation of the binary symbol $1$ is called an
\emph{event}. The event \emph{count} is the number of $0$'s between successive $1$s.
Counts $n$ are i.i.d. random variables drawn from an \emph{interevent distribution} $F(n)$, $n
\geq 0$. We restrict ourselves to \textit{persistent} renewal processes, such
that the probability distribution function is normalized: $\sum_{n=0}^{\infty}
F(n) = 1$. This translates into the processes being ergodic and stationary.
We also define the \emph{survival function} by $w(n) = \sum_{n'=n}^\infty
F(n')$, and the expected interevent count is given by $\mu =
\sum_{n=0}^{\infty} n F(n)$. We assume also that $\mu<\infty$. It is
straightforward to check that $\sum_{n=0}^{\infty} w(n) = \mu + 1$.

Note the dual use of $\mu$. On the one hand, it denotes the measure over sequences and, since it determines probabilities, it appears in names for
informational quantities. On the other, it is a commonplace in renewal process
theory that denotes mean rates. Fortunately, context easily distinguishes the meaning through the very different uses.

\subsection{Process Unpredictability}
\label{sec:Information}

The information or uncertainty in a process is often defined as the Shannon
entropy $H[\MS_0]$ of a single symbol $\MS_0$ \cite{Cove06a}:
\begin{align}
H[\MS_0] = - \sum_{\ms \in \ProcessAlphabet} \Prob(\MS_0=\ms) \log_2 \Prob(\MS_0=\ms)
  ~.
\label{eq:SingleSymbolEntropy}
\end{align}
However, since we are interested in general complex processes---those with
arbitrary dependence structure---we employ the \emph{block entropy} to
monitor information in long sequences:
\begin{align*}
H(L) & = H[\MS_{0:L}] \\
     & = - \sum_{w \in \ProcessAlphabet^L} \Prob(\MS_{0:L}=w) \log_2
	 \Prob(\MS_{0:L}=w)
  ~.
\end{align*}
To measure a process's asymptotic per-symbol uncertainty one then uses the
Shannon entropy rate:
\begin{align*}
\hmu = \lim_{L \to \infty} \frac{H(L)}{L}
  ~,
\end{align*}
when the limit exists. (Here and elsewhere, $\mu$ reminds us that information
quantities depend on the process's measure $\mu$ over sequences.) $\hmu$ quantifies
the rate at which a stochastic process generates information. Using standard
informational identities, one sees that the entropy rate is also given by the
conditional entropy:
\begin{align}
\hmu = \lim_{L \to \infty} H[\MS_0|\MS_{-L:0}] ~.
\label{eq:EntropyRateConditional}
\end{align}
This form makes transparent its interpretation as the residual uncertainty in a
measurement given the infinite past. As such, it is often employed as a
measure of a process's degree of unpredictability.

\subsection{Maximally Predictive Models}
\label{sec:ProcStruct}

Forward-time causal states $\SSet^+$ are minimal sufficient statistics for
predicting a process's future \cite{Crut88a,Shal98a}. This follows from their
definition---a \emph{causal state} $\st^+ \in \SSet^+$ is a sets of pasts grouped by the equivalence relation $\sim^+$:
\begin{align}
\ms_{:0} \sim^+ & \ms_{:0}' \nonumber \\
  & \Leftrightarrow
  \Prob (\MS_{0:}|\MS_{:0}=\ms_{:0}) = \Prob(\MS_{0:}|\MS_{:0}=\ms_{:0}')
  ~.
\end{align}
So, $\SSet^+$ is a set of classes---a coarse-graining of the uncountably
infinite set of all pasts. At time $t$, we have the random variable $\St^+_t$
that takes values $\st^+ \in \SSet^+$ and describes the \emph{causal-state
process} $\ldots, \St^+_{-1}, \St^+_0, \St^+_1, \ldots$. $\St^+_t$ is a
partition of pasts $\MS_{:t}$ that, according to the indexing convention, does
not include the present observation $\MS_t$. In addition to the set of pasts
leading to it, a causal state $\st^+_t$ has an associated \emph{future
morph}---the conditional measure $\Measure(\MS_{t:}| \st^+_t)$ of futures that
can be generated from it. Moreover, each state $\st^+_t$ inherits a probability
$\pi(\st^+_t)$ from the process's measure over pasts $\Measure(\MS_{:t})$. The
forward-time \emph{statistical complexity} is defined as the Shannon entropy of
the probability distribution over forward-time causal states \cite{Crut88a}:
\begin{equation}
\Cmu^+ = H[\St^+_0]
  ~.
\end{equation}

A generative model is constructed out of the causal states by endowing the
causal-state process with transitions:
\begin{align*}
T_{\cs\cs'}^{(\ms)} = \Prob(\St_{t+1}^+=\st',\MS_t = \ms|\St_t^+=\st)
  ~,
\end{align*}
that give the probability of generating the next symbol $\ms$ and ending in
the next state $\cs'$, if starting in state $\cs$. (Residing in a state and
generating a symbol do not occur simultaneously. Since symbols are generated
during transitions there is, in effect, a half time-step difference in the
indexes of the random variables $\MS_t$ and $\St^+_t$. We suppress notating
this.) To summarize, a process's \emph{forward-time \eM} is the tuple
$\{\ProcessAlphabet, \SSet^+, \{ T^{(x)} : \ms \in \ProcessAlphabet \} \}$.

For a discrete-time, discrete-alphabet process, the \eM\ is its minimal
unifilar Hidden Markov Model (HMM) \cite{Crut88a,Shal98a}. (For general
background on HMMs see \cite{Paz71a,Rabi86a,Rabi89a}.) Note that the causal
state set can be finite, countable, or uncountable; the latter two cases can
occur even for processes
generated by finite-state HMMs. \emph{Minimality} can be defined by either the smallest number of states or the smallest entropy over states \cite{Shal98a}. \emph{Unifilarity} is a constraint on the
transition matrices $T^{(x)}$ such that the next state $\st'$ is determined by
knowing the current state $\st$ and the next symbol $x$.
That is, if the transition exists, then $\Prob(\St_{t+1}^+|\MS_t =
\ms,\St_t^+=\st)$ has support on a single causal state.

While the \eM\ is a process's minimal, maximally predictive model, there can be
alternative HMMs that are as predictive, but are not minimal. We refer to the
maximally predictive property by referring to the \eM\ and these alternatives
as \emph{prescient}. The state and transition structure of a prescient model
allow one to immediately calculate the entropy rate $\hmu$, for example.
More generally, any statistic that gives the same (optimal) level of predictability, we call a \emph{prescient statistic}.

A similar equivalence relation can be applied to find minimal sufficient
statistics for retrodiction \cite{Crut08a}. Futures are grouped together if
they have equivalent conditional probability distributions over pasts:
\begin{align}
\ms_{0:} \sim^- & \ms_{0:}' \nonumber \\
  & \Leftrightarrow
  \Prob(\MS_{:0}|\MS_{0:}=\ms_{0:}) = \Prob(\MS_{:0}|\MS_{0:}=\ms_{0:}')
  ~.
\end{align}
A cluster of futures---a \emph{reverse-time causal state}---defined by $\sim^-$
is denoted $\st^-\in\SSet^-$. Again, each $\st^-$ inherits a probability
$\pi(\st^-)$ from the measure over futures $\Measure(\MS_{0:})$. And, the
\emph{reverse-time statistical complexity} is the Shannon entropy of the
probability distribution over reverse-time causal states:
\begin{equation}
\Cmu^- = H[\St^-_0]
  ~.
\end{equation}
In general, the forward and reverse-time statistical complexities are not equal
\cite{Crut08a, Elli11a}. That is, different amounts of information must be
stored from the past (future) to predict (retrodict). Their difference $\Xi =
C_{\mu}^+-C_{\mu}^-$ is a process's \emph{causal irreversibility} and it
reflects this statistical asymmetry.

Since we work with stationary processes in the following the time origin is
arbitrary and so we drop the time index $t$ when it is unnecessary.

\subsection{Information Measures for Processes}
\label{sec:Measures}

Shannon's various information quantities---entropy, conditional entropy, mutual
information, and the like---when applied to time series are functions of the
joint distributions $\Prob(\MS_{0:L})$. Importantly, they define an algebra of
information measures for a given set of random variables \cite{Yeun08a}. Reference
\cite{Jame11a} used this to show that the past and future partition the
single-measurement entropy $H(\MS_0)$ into several distinct measure-theoretic
\emph{atoms}. These include the \emph{ephemeral information}:
\begin{align}
\rmu = H[\MS_0|\MS_{:0},\MS_{1:}] ~,
\end{align}
which measures the uncertainty of the present knowing the past and future;
the \emph{bound information}:
\begin{align}
\bmu = I[\MS_0;\MS_{1:}|\MS_{:0}] ~,
\end{align}
which is the mutual information shared between present and future conditioned
on past; and the \emph{enigmatic information}:
\begin{align}
\qmu = I[\MS_0;\MS_{:0};\MS_{1:}] ~,
\end{align}
which is the three-way mutual information between past, present, and future.
Multi-way mutual informations are sometimes referred to as
\emph{co-informations} \cite{Bell03a,Jame11a} and, compared
to Shannon entropies and two-way mutual information, can have
counterintuitive properties, such as being negative.

For a stationary time series, the bound information is also the
shared information between present and past conditioned on the future:
\begin{align}
\bmu = I[\MS_0;\MS_{:0}|\MS_{1:}].
\end{align}
One can also consider the amount of predictable information not captured by the present:
\begin{align}
\sigmamu = I[\MS_{:0};\MS_{1:}|\MS_0]
  ~.
\end{align}
This is called the \emph{elusive information} \cite{Moha14a}. It measures the amount of
past-future correlation not contained in the present. It is nonzero if the
process \emph{necessarily} has hidden states and is therefore quite sensitive to how the state
space is observed or coarse grained.

The maximum amount of information in the future predictable from the past
(or vice versa) is the \emph{excess entropy}:
\begin{align*}
\EE & = I[\MS_{:0};\MS_{0:}] ~.
\end{align*}
It is symmetric in time and a lower bound on the stored
informations $\Cmu^+$ and $\Cmu^-$.
It is directly given by the information atoms above:
\begin{align}
\EE & = \bmu + \sigmamu + \qmu ~.
\label{eq:EE1}
\end{align}
The process's Shannon entropy rate $\hmu$---recall the form of Eq.
(\ref{eq:EntropyRateConditional})---can also be written as a sum of atoms:
\begin{align*}
\hmu & = H[\MS_0|\MS_{:0}] \\
     & = \rmu + \bmu
	 ~.
\end{align*}
Thus, a portion of the information ($\hmu$) a process spontaneously generates is thrown away ($\rmu$) and a portion is actively stored ($\bmu$). Putting
these observations together gives the information architecture of a single
measurement (Eq. (\ref{eq:SingleSymbolEntropy})):
\begin{equation}
H[\MS_0] = \qmu + 2\bmu + \rmu ~.
\label{eq:H01}
\end{equation}
These identities can be used to determine $\rmu$, $\qmu$, and $\EE$
from $H[X_0]$, $\bmu$, and $\sigmamu$, for example.

We have a particular interest in when $\Cmu$ and $\EE$ are
infinite and so will investigate finite-time variants of causal states and finite-time estimates of statistical complexity and $\EE$. For example, the latter is given by:
\begin{equation}
\Ipred(M,N) = I[\MS_{-M:0};\MS_{0:N}]
  ~.
\end{equation}
If $\EE$ is finite, then $\EE = \lim_{M,N\rightarrow\infty} \Ipred(M,N)$. When
$\EE$ is infinite, then the way in which $\Ipred(M,N)$ diverges is one measure
of a process' complexity \cite{Crut89e,Bial00a,Crut01a}. Analogous, finite
past-future $(M,N)$-parametrized equivalence relations lead to finite-time
forward and reverse
causal states and statistical complexities $\Cmu^+(M,N)$ and $\Cmu^-(M,N)$.


\section{Causal Architecture of Renewal Processes}
\label{sec:CS_DTSRP}

\begin{figure}[h!]
\includegraphics[width=0.2\textwidth]{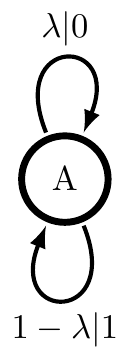}
\includegraphics[width=0.61\textwidth]{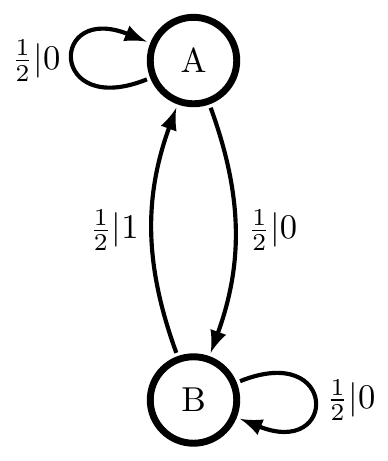}
\caption{(Left) Minimal generative model for the Poisson process with rate
  $\lambda$. (Right) A generator for the Simple Nonunifilar Source (SNS).
  Both generate a stationary renewal process.
  Transition labels $p|s$ denote probability $p$ of taking a
  transition and emitting symbol $s$.
  }
\label{fig:SimpleExamples}
\end{figure}

It will be helpful pedagogically to anchor our theory in the contrast between
two different, but still simple, renewal processes. One is the familiar
``memoryless'' Poisson process with rate $\lambda$. Its HMM generator, a biased
coin, is shown at the left of Fig.~\ref{fig:SimpleExamples}. It has an
interevent count distribution $F(n) = (1-\lambda)\lambda^n$; a distribution
with unbounded support. However, we notice in Fig.~\ref{fig:SimpleExamples}
that it is a unifilar model with a minimal number of states. So, in fact, this
one-state machine \textit{is} the \eM\ of a Poisson process. The rate at which
it generates information is given by the entropy rate: $\hmu = H(\lambda)$ bits
per output symbol. (Here, $H(p)$ is the binary entropy function.) It also has a
vanishing statistical complexity $\Cmu^+ = 0$ and so stores no historical
information.

The second example is the \emph{Simple Nonunifilar Source} (SNS)
\cite{Crut92c}; an HMM generator for which is shown on the right of
Fig.~\ref{fig:SimpleExamples}. Transitions from state $B$ are
unifilar, but transitions from state $A$ are not. In fact, a
little reflection shows that the time series produced by the SNS
is a discrete-time renewal process. Once we observe the ``event''
$\ms_t = 1$, we know the internal model state to be $\st_{t+1} = A$, so successive interevent counts are completely uncorrelated.

The SNS generator is not an \eM\ and, moreover, it cannot be used to calculate
the process's information per output symbol (entropy rate). If we can only see $0$'s and $1$'s, we will usually be uncertain as to whether we are in state $A$ or state $B$, so this generative model is not maximally predictive. How can we
calculate this basic quantity? And, if we cannot use the two-state generator,
how many states are required and what is their transition dynamic? The
following uses computational mechanics to answer these and a number of related questions. To aid readability, though, we sequester most all of the detailed calculations and proofs in App. \ref{app:CausalArchitecture}.

We start with a simple Lemma that follows directly from the definitions of a
renewal process and the causal states. It allows us to introduce notation that
simplifies the development.

{\Lem The count since last event is a prescient statistic of a discrete-time stationary renewal process.
\label{lem:A}
}

That is, if we remember only the number of counts since the last event and
nothing prior, we can predict the future as well as if we had memorized the
entire past. Specifically, a \emph{prescient state} $\AlternateState$ is a function
of the past such that:
\begin{align*}
H[\Future|\Past] = H[\Future|\AlternateState]
  ~.
\end{align*}

Causal states can be written as unions of prescient states \cite{Shal98a}. We start with a definition that helps to characterize the converse; i.e., when the
prescient states of Lemma \ref{lem:A} are also causal states.

To ground our intuition, recall that Poisson processes are ``memoryless''. This
may seem counterintuitive, if viewed from a parameter estimation point of view.
After all, if observing longer pasts, one makes better and better estimates of
the Poisson rate. However, finite data fluctuations in estimating model
parameters are irrelevant to the present mathematical setting \textit{unless}
the parameters are themselves random variables, as in Ref. \cite{Bial00a}.
This is not our setting here: the parameters are fixed. In fact, we restrict
ourselves to studying ergodic processes, in which the conditional probability
distributions of futures given pasts of a Poisson process are
independent of the past.

We therefore expect the prescient states in Lemma \ref{lem:A} to fail to be
causal states precisely when the interevent distribution is similar to that of
a Poisson renewal process. This intuition is made precise by Def.
\ref{lem:MultipoissonInterevent}.

\begin{figure*}[htb]
\begin{floatrow}
\floatbox{figure}[.22\textwidth][\FBheight][t]
	{\caption{Not eventually $\Delta$-Poisson with unbounded-support interevent distribution.}
	\label{fig:Generic_eM1}
	}
	{\includegraphics[width=.24\textwidth]{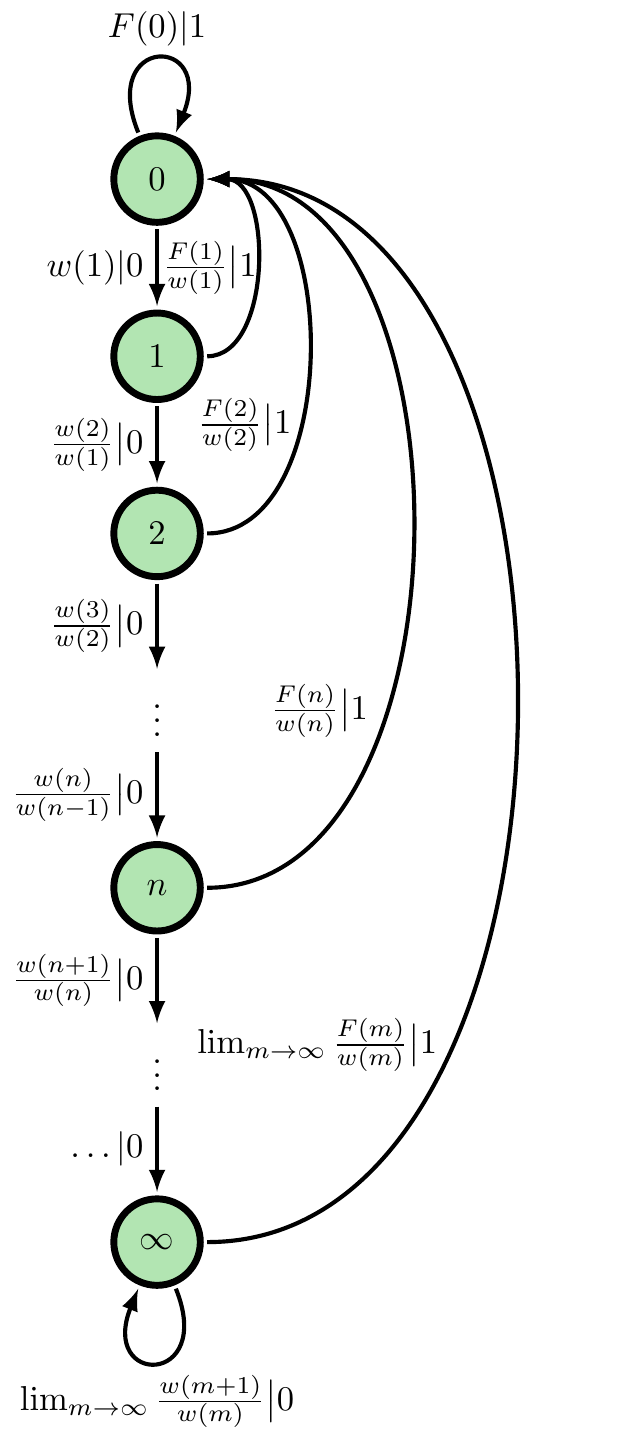}}
\floatbox{figure}[.22\textwidth][\FBheight][t]
	{\caption{Not eventually $\Delta$-Poisson with bounded-support interevent distribution.}
	\label{fig:Generic_eM2}
	}
	{\includegraphics[width=.24\textwidth]{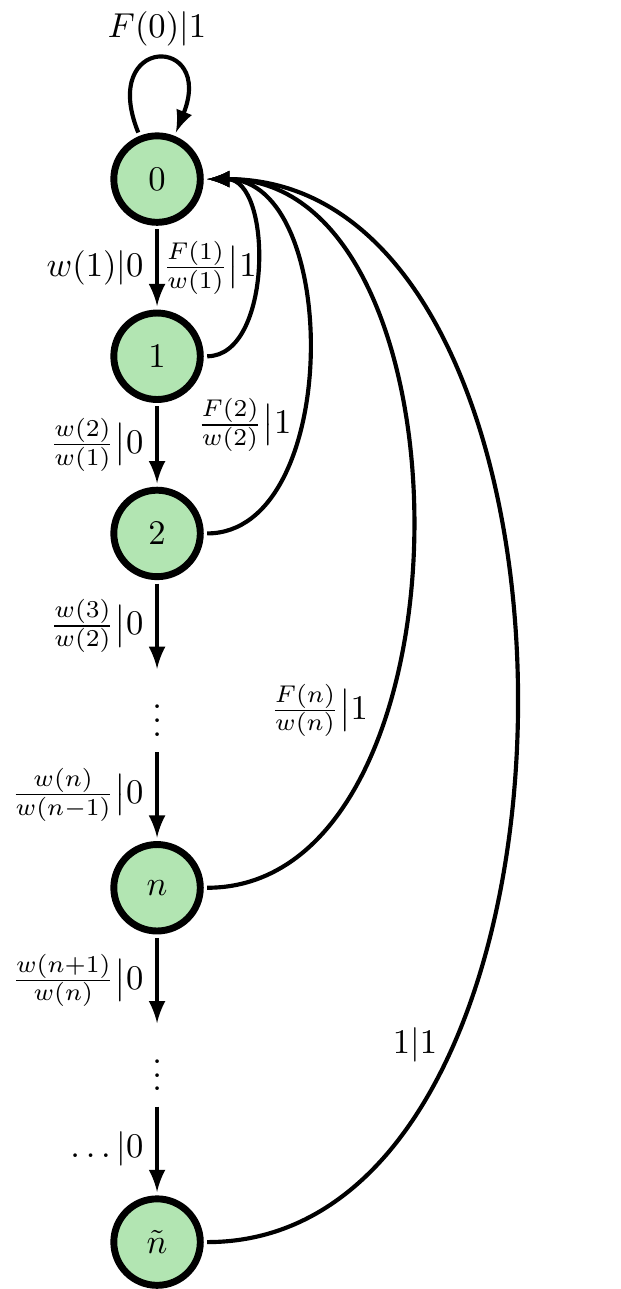}}
	\hspace*{0.5cm}
\floatbox{figure}[.22\textwidth][\FBheight][t]
	{\caption{Eventually $\Delta$-Poisson with characteristic $(\tilde{n},\Delta=1)$
in Def. \ref{lem:MultipoissonInterevent}.}
	\label{fig:Generic_eM3}
	}
	{\includegraphics[width=.24\textwidth]{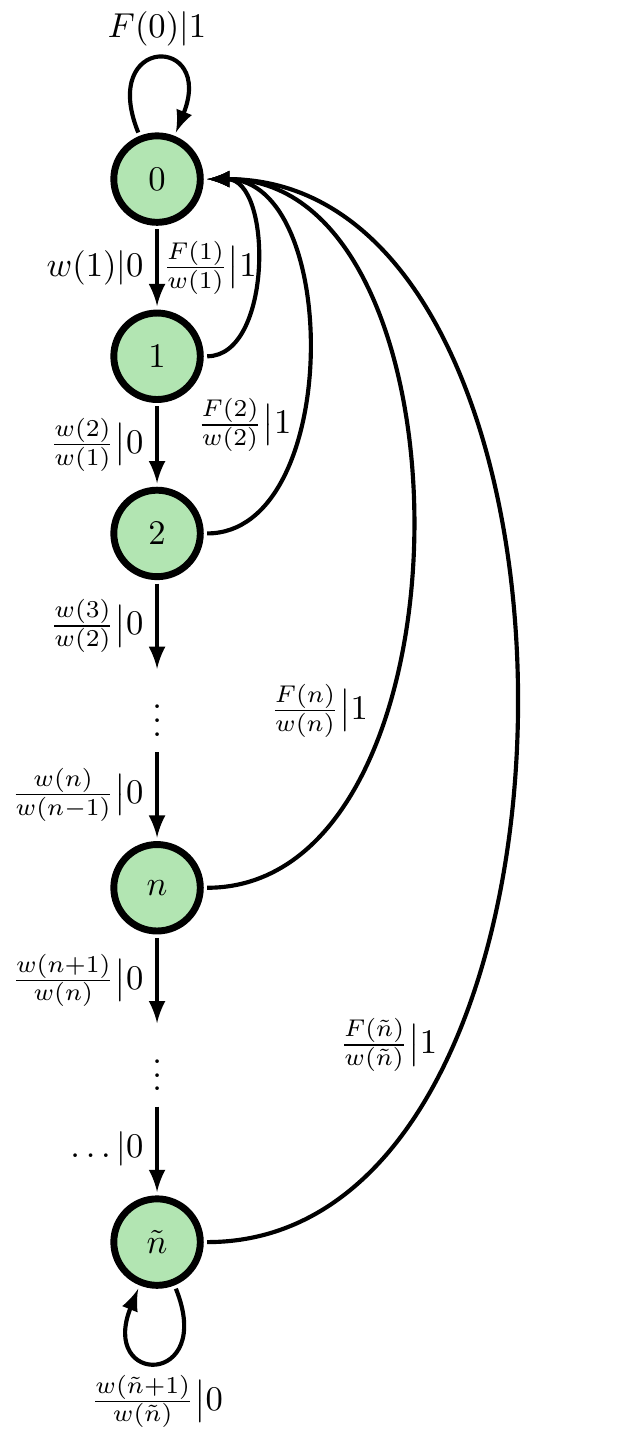}}
\floatbox{figure}[.22\textwidth][\FBheight][t]
	{\caption{Eventually $\Delta$-Poisson with characteristic $(\tilde{n},\Delta>1)$
 	in Def. \ref{lem:MultipoissonInterevent}.}
	\label{fig:Generic_eM4}
	}
	{\includegraphics[width=.24\textwidth]{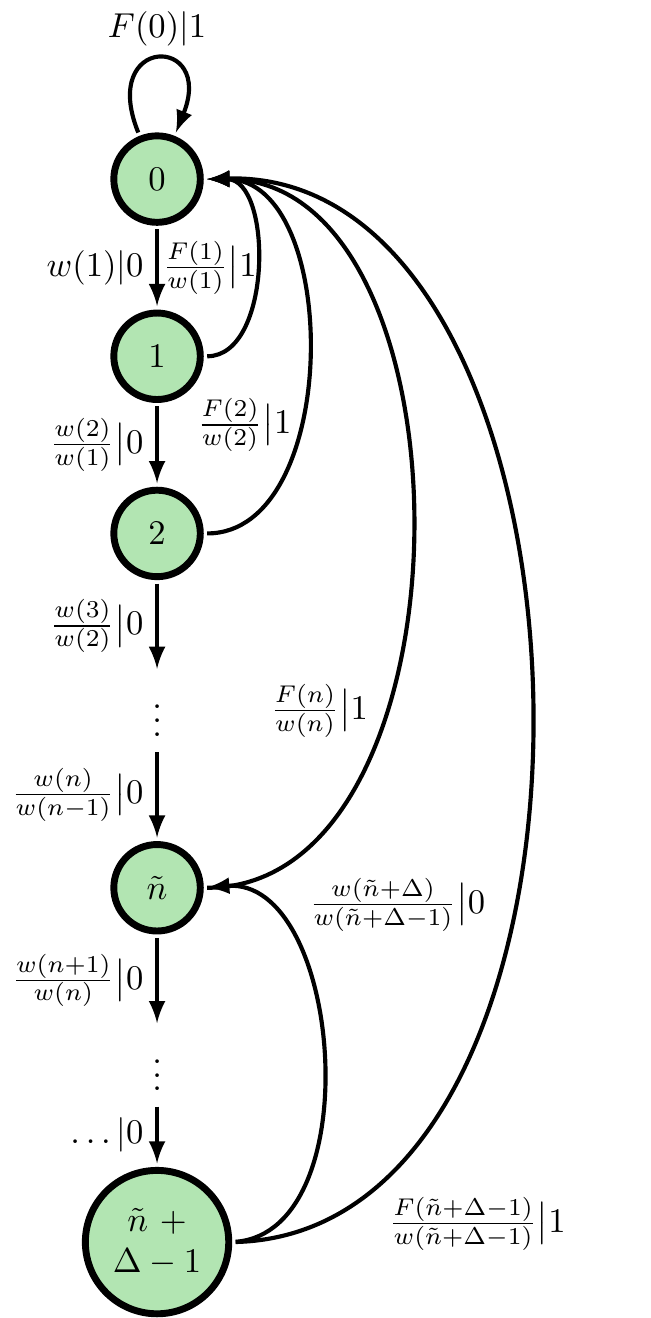}}
\end{floatrow}
\vspace{0.5 cm}
{Possible \eM\ architectures for discrete-time stationary renewal processes.}
\end{figure*}

{\Def A \emph{$\Delta$-Poisson} process has an interevent distribution
\begin{align*}
F(n) = F(n \!\!\!\! \mod \Delta) ~ \lambda^{\lfloor n/\Delta \rfloor}
  ~,
\end{align*}
for all $n$ and some $\lambda>0$.  If this statement holds for multiple $\Delta\geq 1$, then we choose the smallest possible $\Delta$.}

{\Def
A $(\tilde{n},\Delta)$
\emph{eventually $\Delta$-Poisson} process has an interevent distribution that
is $\Delta$-Poisson for all $n\geq \tilde{n}$:
\begin{align*}
F(\tilde{n}+k\Delta+m) = \lambda^k F(\tilde{n}+m)
  ~,
\end{align*}
for all $0 \leq m < \Delta$, for all $k\geq 0$, and for some $\lambda > 0$.
If this statement holds for multiple $\Delta \geq 1$ and multiple $\tilde{n}$, then we choose the smallest possible $\Delta$ and the smallest possible $\tilde{n}$.
\label{lem:MultipoissonInterevent}
}

Thus, a Poisson process is a $\Delta$-Poisson process with $\Delta = 1$ and an
eventually $\Delta$-Poisson process with $\Delta=1$ and $\tilde{n}=0$.
Moreover, we will now show that at some finite $\tilde{n}$, any renewal process
is either (i) Poisson, if $\Delta=1$, or (ii) a combination of several Poisson
processes, if $\Delta>1$.

Why identify new classes of renewal process? In short, renewal processes that
are similar to, but not the same as, the Poisson process do not have an
infinite number of causal states. The particular condition for when they do not
is given by the eventually $\Delta$-Poisson definition. Notably, this new class
is what emerged, rather unexpectedly, by applying the causal-state equivalence
relation $\sim^+$ to renewal processes. The resulting insight is that general
renewal processes, after some number of counts (the ``eventually'' part) and
after some coarse-graining of counts (the $\Delta$ part), behave like a Poisson
process.

With these definitions in hand, we can proceed to identify the causal
architecture of discrete-time stationary renewal processes.

{\The (a) The forward-time causal states of a discrete-time stationary renewal
process that is not eventually $\Delta$-Poisson are groupings of pasts with the same count
since last event. (b) The forward-time causal states of a discrete-time
eventually $\Delta$-Poisson stationary renewal process are groupings of pasts with the same
count since last event up until $\tilde{n}$ and pasts whose count $n$ since last event are in the same equivalence class as $\tilde{n}$ modulo $\Delta$.
\label{thm:1}
}

The Poisson process, as an eventually $\Delta$-Poisson with $\tilde{n}=0$ and
$\Delta=1$, is represented by the one-state \eM\ despite the unbounded support
of its interevent count distribution.  Unlike most processes, the Poisson
process' \eM\ is the same as its generative model shown in
Fig.~\ref{fig:SimpleExamples}(left).

The SNS, on the other hand, has an interevent count distribution that is not
eventually $\Delta$-Poisson. According to Thm.~\ref{thm:1}, then, the SNS has a
countable infinity of causal states despite its simple two-state generative
model in Fig.~\ref{fig:SimpleExamples}(left). Compare Fig.
\ref{fig:Generic_eM1}. Each causal state corresponds to a different probability
distribution over the internal states $A$ and $B$. These internal state
distributions are the \emph{mixed states} of Ref. \cite{Crut08b}. Observing
more $0$'s, one becomes increasingly convinced that the internal state is
$B$. For maximal predictive power, however, we must track the probability
that the process is still in state $A$. Both Fig.~\ref{fig:Generic_eM1} and
Fig.~\ref{fig:SimpleExamples}(left)~ are ``minimally complex'' models of the
same process, but with different definitions of model complexity. We return to
this point in Sec. \ref{sec:SNS}.

Appendix \ref{app:CausalArchitecture} makes the statements in Thm.~\ref{thm:1}
precisely. The main result is that causal states are sensitive to two features:
(i) eventually
$\Delta$-Poisson structure in the interevent distribution and (ii) the boundedness
of $F(n)$'s support. If the support is bounded, then there are a finite number
of causal states rather than a countable infinity of causal states.  Similarly,
if $F(n)$ has $\Delta$-Poisson tails, then there are a finite number of causal
states despite the support of $F(n)$ having no bound.  Nonetheless, one can say
that the generic discrete-time stationary renewal process has a countable
infinity of causal states.

Finding the probability distribution over these causal states is
straightforwardly related to the survival-time distribution $w(n)$ and the mean
interevent interval $\mu$, since the probability of observing at least $n$ counts
since last event is $w(n)$. Hence, the probability of seeing $n$ counts since
the last event is simply the normalized survival function $w(n) / (\mu + 1)$. Appendix \ref{app:CausalArchitecture} derives the statistical complexity using this and Theorem \ref{thm:1}. The resulting formulae are given
in Table \ref{tab:MeasuresFormulae} for the various cases.  

As described in Sec. \ref{sec:Background}, we can also endow the causal
state space with a transition dynamic in order to construct the renewal process
\eM---the process's minimal unifilar hidden Markov model. The transition
dynamic is sensitive to $F(n)$'s support and not only its boundedness. For
instance, the probability of observing an event given that it has been $n$
counts since the last event is $F(n) / w(n)$. For the generic discrete-time
renewal process this is exactly the transition probability from causal state
$n$ to causal state $0$. If $F(n)=0$, then there is no probability of
transition from $\st=n$ to $\st=0$. See
App. \ref{app:CausalArchitecture} for details.

\setlength\extrarowheight{5pt}

\begin{table*}[ht] 
\centering
\begin{tabular}{l l}
\hline\hline
\bf Quantity & \bf Expression \\
\hline
$\Cmu^+ = H[\St^+]$
  & $-\sum_{n=0}^{\infty} \frac{w(n)}{\MIET}\log_2 \frac{w(n)}{\MIET}$
  \hspace{2.6in} Not eventually $\Delta$-Poisson \\[5pt]
  & $- \sum_{n=0}^{\tilde{n}-1} \frac{w(n)}{\MIET}\log_2\frac{w(n)}{\MIET}
  - \sum_{m=0}^{\Delta-1}
  \frac{\sum_{k=0}^{\infty} w(\tilde{n}+k\Delta+m)}{\MIET}
  \log_2 \frac{\sum_{k=0}^{\infty} w(\tilde{n}+k\Delta+m)}{\MIET}$
  Eventually $\Delta$-Poisson \\[5pt]
\hline
$\EE = I[\MS_{:0};\MS_{0:}]$ & $-2 \sum_{n=0}^{\infty}
  \frac{w(n)}{\MIET}\log_2 \frac{w(n)}{\MIET}
  + \sum_{n=0}^{\infty} (n+1) \frac{F(n)}{\MIET} \log_2 \frac{F(n)}{\MIET}$
  \\[5pt]
\hline
$\hmu = H[\MS_0|\MS_{:0}]$ & $-\MIETI \sum_{n=0}^{\infty} F(n)\log_2 F(n)$
  \\[5pt]
\hline
$\bmu = I[\MS_{1:};\MS_0|\MS_{:0}]$
  & ~ $\MIETI \big\{ \sum_{n=0}^{\infty}
  (n+1)F(n)\log_2 F(n) - \sum_{m,n=0}^{\infty} g(m,n)\log_2 g(m,n) \big\}$
  \\[5pt]
\hline
$\sigmamu = I[\MS_{1:};\MS_{:0}|\MS_0]$
  & ~ $\MIETI \big\{ \mu \log_2 \mu + \sum_{n=0}^{\infty} n F(n)\log_2 F(n)
  - 2 \sum_{n=0}^{\infty} w(n)\log_2 w(n) \big\}$ \\[5pt]
\hline
$\qmu = I[\MS_{1:};\MS_0;\MS_{:0}]$
  & ~ $\MIETI \big\{ \sum_{m,n=0}^{\infty} g(m,n)\log_2 g(m,n)
 - \sum_{n=0}^{\infty} w(n)\log_2 w(n)
 + (\MIET) \log_2 (\MIET) - \mu \log_2 \mu \big\}$ \\[5pt]
 \hline
$\rmu = H[\MS_0|\MS_{1:},\MS_{:0}]$
  & ~ $\MIETI \big\{ \sum_{m,n=0}^{\infty} g(m,n)\log_2 g(m,n)
  - \sum_{n=0}^{\infty} (n+2)F(n) \log_2 F(n) \big\}$ \\[5pt]
\hline
$H_0 = H[\MS_0]$ & $-\MIETI \log_2 \MIETI
  - \big( 1-\MIETI \big) \log_2 \big( 1-\MIETI \big)$ \\[5pt]
\hline\hline
\end{tabular} 
\caption{Structural measures and information architecture of a stationary renewal
  process with interevent counts drawn from the distribution $F(n)$, $n\geq 0$,
  survival count distribution $w(n) = \sum_{m=n}^{\infty} F(m)$, and mean
  interevent count $\mu = \sum_{n=0}^{\infty} nF(n) <\infty$. The function
  $g(m,n)$ is defined by $g(m,n) = F(m+n+1)+F(m)F(n)$. Cases are needed for
  $\Cmu$ but not other quantities, such as block entropy and information
  architecture quantities, since the latter can be calculated just as well from
  prescient machines.  The quantities $\chi$ (crypticity) and $\Ipred(M,N) =
  I[\MS_{-M:0};\MS_{0:N}]$ are no less interesting than the others given here,
  but their expressions are not compact; see App.
  \ref{app:InformationAnatomy}.
  }
\label{tab:MeasuresFormulae}
\end{table*}

%

Figures \ref{fig:Generic_eM1}-\ref{fig:Generic_eM4} display the
causal state architectures, depicted as state-transition diagrams, for the \eMs\ in the various cases delineated. Figure
\ref{fig:Generic_eM1} is the \eM\ of a generic renewal process whose interevent
interval can be arbitrarily large and whose interevent distribution never has
exponential tails. Figure \ref{fig:Generic_eM2} is the \eM\ of a renewal
process whose interevent distribution never has exponential tails but cannot have arbitrarily large interevent counts. The \eM\ in Fig.
\ref{fig:Generic_eM3} looks quite similar to the \eM\ in Fig.
\ref{fig:Generic_eM2}, but it has an additional transition that connects the
last state $\tilde{n}$ to itself. This added transition changes our
structural interpretation of the process.  Interevent counts can be arbitrarily large for
this \eM\ but past an interevent count of $\tilde{n}$, the interevent
distribution is exponential. Finally, the \eM\ in Fig. \ref{fig:Generic_eM4}
represents an eventually $\Delta$-Poisson process with $\Delta>1$ whose
structure is conceptually most similar to that of the \eM\ in Fig.
\ref{fig:Generic_eM3}. (See Def. \ref{lem:MultipoissonInterevent} for the
precise version of that statement.)  If our renewal process disallows seeing
interevent counts of a particular length $L$, then this will be apparent from
the \eM\ since there will be no transition between the causal state
corresponding to an interevent count of $L$ and causal state $0$.

As described in Sec. \ref{sec:Background}, we can analytically characterize
a process' information architecture far better once we characterize its
statistical structure in reverse time.

{\Lem Groupings of futures with the same counts to next event are reverse-time prescient statistics for discrete-time stationary renewal processes.
\label{cor:ReversePrescientCounts}
}

{\The (a) The reverse-time causal states of a discrete-time stationary renewal
process that is not eventually $\Delta$-Poisson are groupings of futures with the same count
to next event.
(b) The reverse-time causal states of a discrete-time eventually $\Delta$-Poisson stationary renewal
process are groupings of futures with the same count to next event up until
$\tilde{n}$ plus groupings of futures whose count since last event $n$ are in
the same equivalence class as $\tilde{n}$ modulo $\Delta$.
\label{cor:1}
}

As a result, in reverse time a stationary renewal process is effectively the same stationary renewal process---counts between events are still independently drawn from $F(n)$. Thus, the causal irreversibility vanishes: $\CI = 0$.

Moreover, these results taken together indicate that we can straightforwardly
build a renewal process's \emph{bidirectional machine} from these forward and
reverse-time causal states, as described in Refs.
\cite{Crut08a,Crut08b,Elli11a}.  Additional properties can then be deduced from
the bidirectional machine, but we leave this for the future.

\section{Information Architecture of Renewal Processes}
\label{sec:EM}

As Sec. \ref{sec:Background} described, many quantities that capture a
process's predictability and randomness can be calculated from knowing the
block entropy function $H(L)$. Often, the block entropy is estimated by
generating samples of a process and estimating the entropy of a trajectory
distribution.  This method has the obvious disadvantage that at large $L$,
there are $|\MeasAlphabet|^L$ possible trajectories and $|\MeasAlphabet|^{\hmu
L}$ typical trajectories. And so, one easily runs into the problem of
severe undersampling, previously referred to as the curse of dimensionality.
This matters most when the underlying process has long-range temporal correlations.

Nor can one calculate the block entropy and other such information measures
exactly from generative models that are not maximally predictive (prescient). Then, the model states do not shield the past from the future. For instance,
as noted above, one cannot calculate the SNS's entropy rate from its simple
two-state generative HMM. The entropy of the next symbol given the generative
model's current state ($A$ or $B$) actually \textit{underestimates} the true
entropy rate by assuming that we can almost always precisely determine the
underlying model state from the past. For a sense of the fundamental challenge,
see Refs. \cite{Blac57b, Birc61a}.

However, we \textit{can} calculate the block entropy and various other
information measures in closed-form from a maximally predictive model. In other
words, finding an \eM\ allows one to avoid the curse of dimensionality
inherently involved in calculating the entropy rate, excess entropy, or the other information measures discussed here.

Figure~\ref{fig:setup} summarized the above points. This section makes good on
the procedure outlined there by providing analytic formulae for various
information measures of renewal processes. The formulae for the entropy rate of
a renewal process is already well known, but all others are new.

Prescient HMMs built from the
prescient statistics of Lemma \ref{lem:A} are maximally predictive models, and correspond to the unifilar Hidden Markov Model shown in Fig.
\ref{fig:Generic_eM1}. The prescient machines make no distinction between
eventually $\Delta$-Poisson renewal processes and one that is not, but they do
contain information about the support of $F(n)$ through their transition
dynamics. (See App. \ref{app:CausalArchitecture}.)
Appendix \ref{app:InformationAnatomy} describes how a prescient machine can be
used to calculate all information architecture quantities---$\rmu$, $\bmu$,
$\sigmamu$, $\qmu$, and the more familiar Shannon entropy rate $\hmu$ and
excess entropy $\EE$. A general strategy for calculating these quantities, as
described in Sec. \ref{sec:Background} and Refs. \cite{Jame11a, Marz14a}, is to
calculate $\bmu$, $\hmu$, $\EE$, and $H[\MS_0]$, and then to derive the other
quantities using the information-theoretic identities given in Sec.
\ref{sec:Background}.

Table \ref{tab:MeasuresFormulae} gives the results of these calculations.  It
helps one's interpretation to consider two base cases.  For a Poisson
process, we gain no predictive power by remembering specific pasts, and we would expect the statistical
complexity, excess entropy, and bound information rate to vanish. The entropy
rate and ephemeral information, though, are nonzero.  One can check that this
is, indeed, the case.  For a periodic process with period $T$, in contrast, one
can check that $\mu+1=T$, since the period is the length of the string of $0$'s
(mean interevent interval $\mu$) concatenated with the subsequent event $\ms
= 1$. The
statistical complexity and excess entropy of this process are $\log_2 T$ and
the entropy rate is $\hmu = 0$, as expected.

Calculating the predictable information $\Ipred(M,N)$ requires identifying
finite-time prescient statistics, since the predictable information is the
mutual information between forward-time causal states over pasts of length $M$
and reverse-time causal states over futures of length $N$. Such finite-time
prescient statistics are identified in Corollary \ref{cor:2}, below, and the
predictable information is derived in App. \ref{app:InformationAnatomy}.  The
final expression is not included in Table \ref{tab:MeasuresFormulae} due to its length.

{\Cor
Forward-time (and reverse-time) finite-time $M$ prescient states of a discrete-time stationary renewal process are the counts from (and to) the next event up until and including $M$.
\label{cor:2}
}

\begin{figure}[h!]
\includegraphics[width=1.0\textwidth]{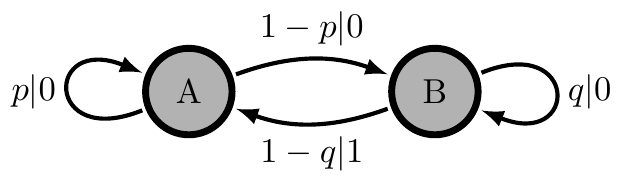}
\includegraphics[width=1.0\textwidth]{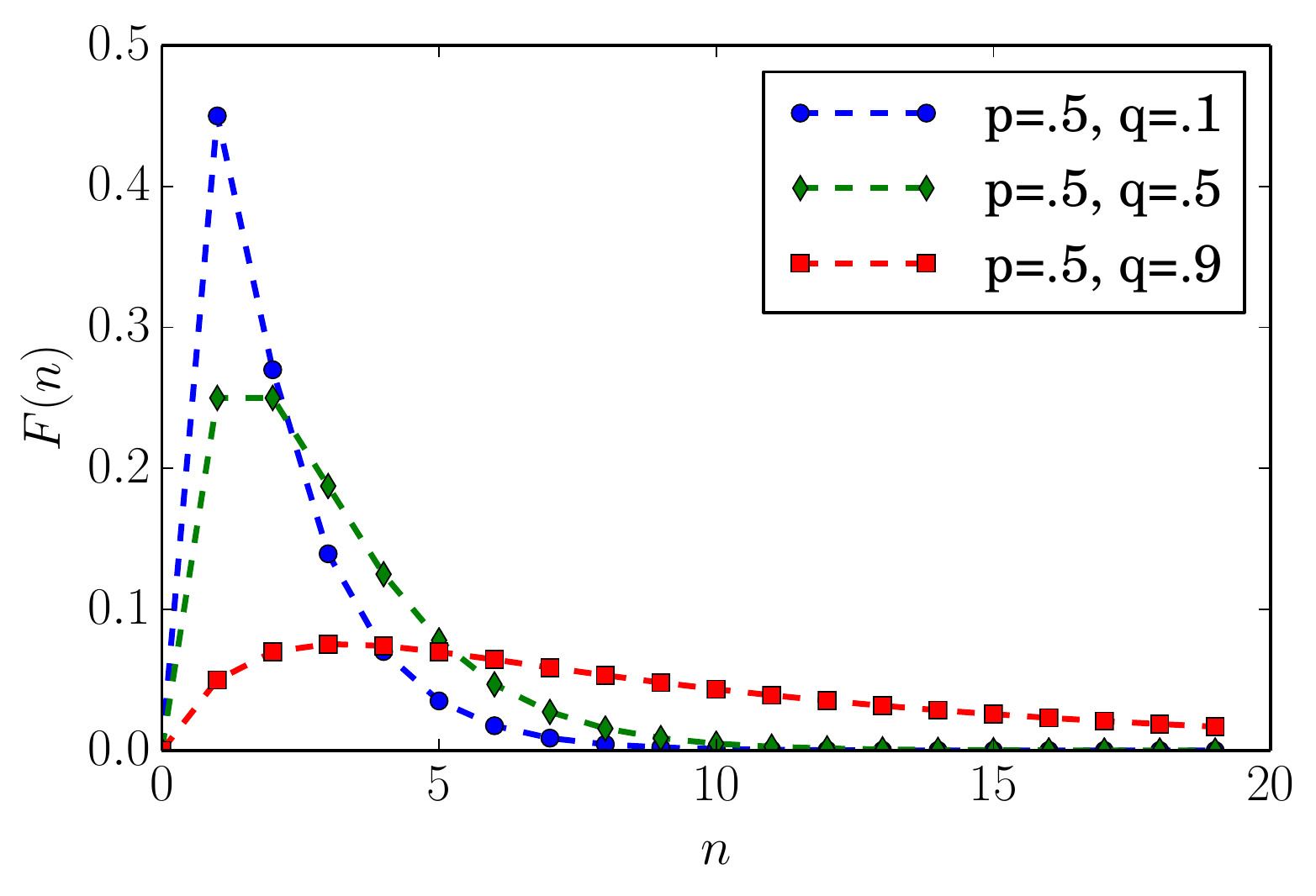}
\caption{(Top) Hidden Markov model for the $(p,q)$ parametrized SNS.
  (Bottom) Example interevent distributions $F(n)$ from
  Eq.~(\ref{eq:FofN_SNS}) for three parameter settings of $(p,q)$.
  }
\label{fig:FofN_SNS}
\end{figure}

All of these quantities can be calculated using a mixed-state presentation, as
described in Ref. \cite{Crut08b}, though the formulae developed there are as yet
unable to describe processes with a countably infinite set of mixed states.
Calculations of finite-time entropy rate estimates using a mixed-state
presentation are consistent with all other results here, though. Purely for
simplicity, we avoid discussing mixed-state presentations.

\section{Nonunifilar HMMs and Renewal Processes}
\label{sec:SNS}

The task of inferring an \eM\ for discrete-time, discrete-alphabet processes is
essentially that of inferring minimal unifilar HMMs; what are sometimes also
called ``probabilistic deterministic'' finite automata. In unifilar HMMs, the
transition to the next hidden state given the previous one and
next emitted symbol is determined. Nonunifilar HMMs are a more general class
of time series models in which the transitions between underlying states given
the next emitted symbol can be stochastic.

\begin{figure*}
\includegraphics[width=\textwidth]{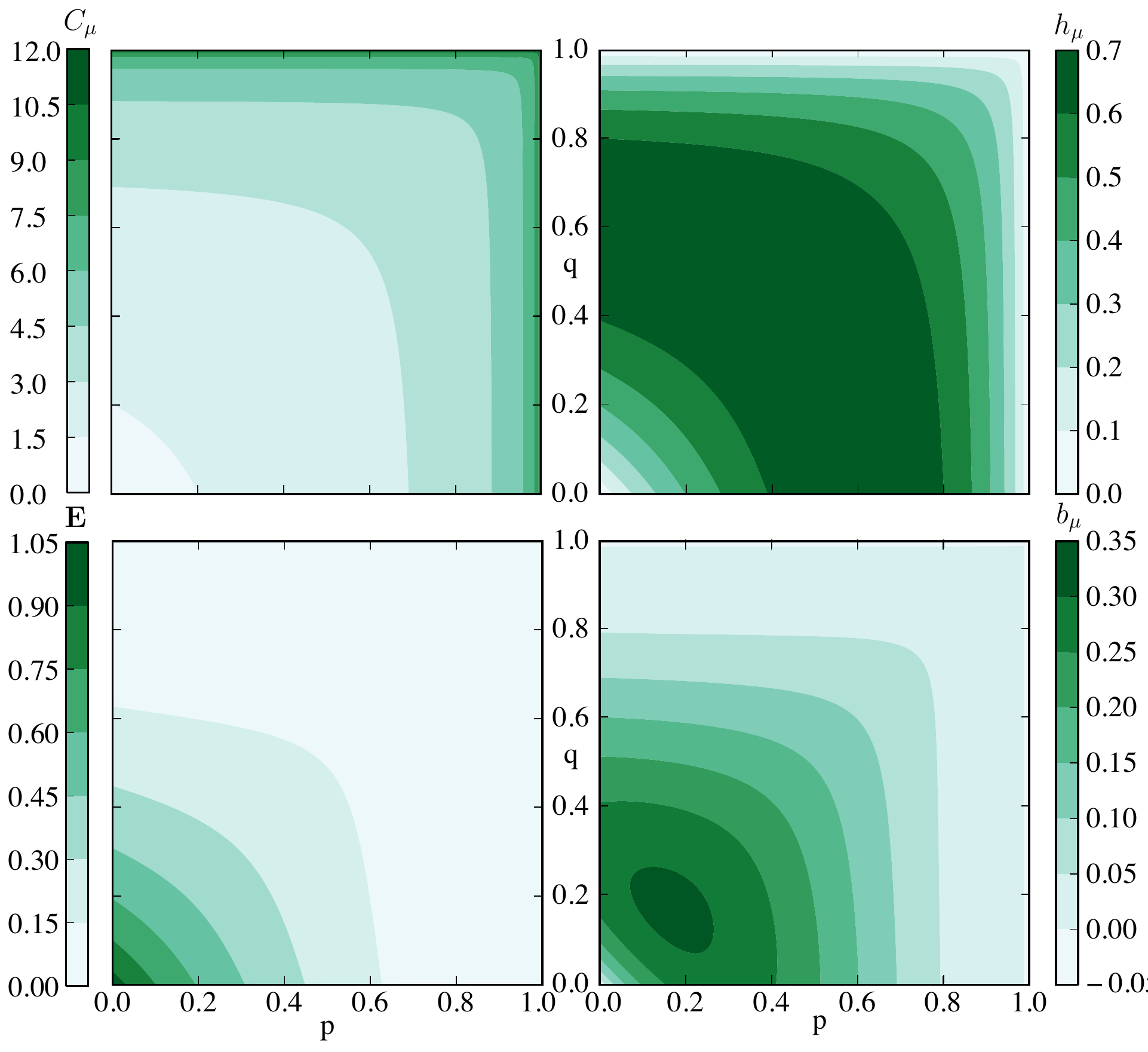}
\caption{Contour plots of various information measures (in bits) as functions of
  SNS parameters $p$ and $q$. (Top left) $\Cmu$, increasing when $F(n)$ has
  slower decay. (Top right) $\hmu$, higher when transition probabilities are
  maximally stochastic. (Bottom left) $\EE$, higher the closer the SNS comes to
  period-$2$. (Bottom right) $\bmu$, highest between the maximally stochastic
  transition probabilities that maximize $\hmu$ and maximally deterministic
  transition probabilities that maximize $\EE$.
  }
\label{fig:SNS_Cmuetc}
\end{figure*}

This simple difference in HMM structure has important consequences for
calculating the predictable information, information architecture, and
statistical complexity of time series generated by nonunifilar HMMs. First,
note that for processes with a finite number of transient and recurrent causal
states, these quantities can be calculated in closed form \cite{Crut13a}.
Second, the autocorrelation function and power spectrum can also be calculated
in closed form for nonunifilar presentations \cite{Riec14a}. Unlike these
cases, though, most of Table \ref{tab:MeasuresFormulae}'s quantities defy
current calculational techniques. As a result, exact calculations of these
prediction-related information measures for even the simplest nonunifilar HMMs
can be surprisingly difficult.

To illustrate this point, we focus our attention on a parametrized version of
the SNS shown in Fig. \ref{fig:FofN_SNS}. As for the original SNS in
Fig.~\ref{fig:SimpleExamples}, transitions from state $B$ are unifilar, but
transitions from state $A$ are not. As noted before, the time series generated
by the parametrized SNS is a discrete-time renewal process with interevent
count distribution:
\begin{align}
F(n) =
  \begin{cases}
  (1-p)(1-q)(p^n-q^n) / (p-q) & p\neq q ~, \\
  (1-p)^2 n p^{n-1} & p=q ~.
  \end{cases}
\label{eq:FofN_SNS}
\end{align}
Figure \ref{fig:FofN_SNS} also shows $F(n)$ at various parameter choices. The
nonunifilar HMM there should be contrasted with the unifilar HMM presentation
of the parametrized SNS which is the \eM\ in Fig.
\ref{fig:Generic_eM1}, with a countable infinity of causal states.

Both parametrized SNS presentations are ``minimally complex'', but according to
different metrics. On the one hand, the nonunifilar presentation is a minimal
\textit{generative} model: No one-state HMM (i.e., biased coin) can produce a
time series with the same statistics. On the other, the unifilar HMM is the
minimal \textit{maximally predictive} model: In order to predict the future as
well as possible given the entire past, we must at least remember how many
$0$'s have been seen since the last $1$. That memory requires a countable
infinity of prescient states. The preferred complexity metric is a matter of taste and desired
implementation,
modulo important concerns regarding
overfitting or ease of inference \cite{Stre13a}. However, if we wish to
calculate the information measures in Table \ref{tab:MeasuresFormulae} as
accurately as possible, finding a maximally predictive model---a unifilar
presentation, that is---is necessary.

\begin{figure}[h!]
\includegraphics[width=1.0\textwidth]{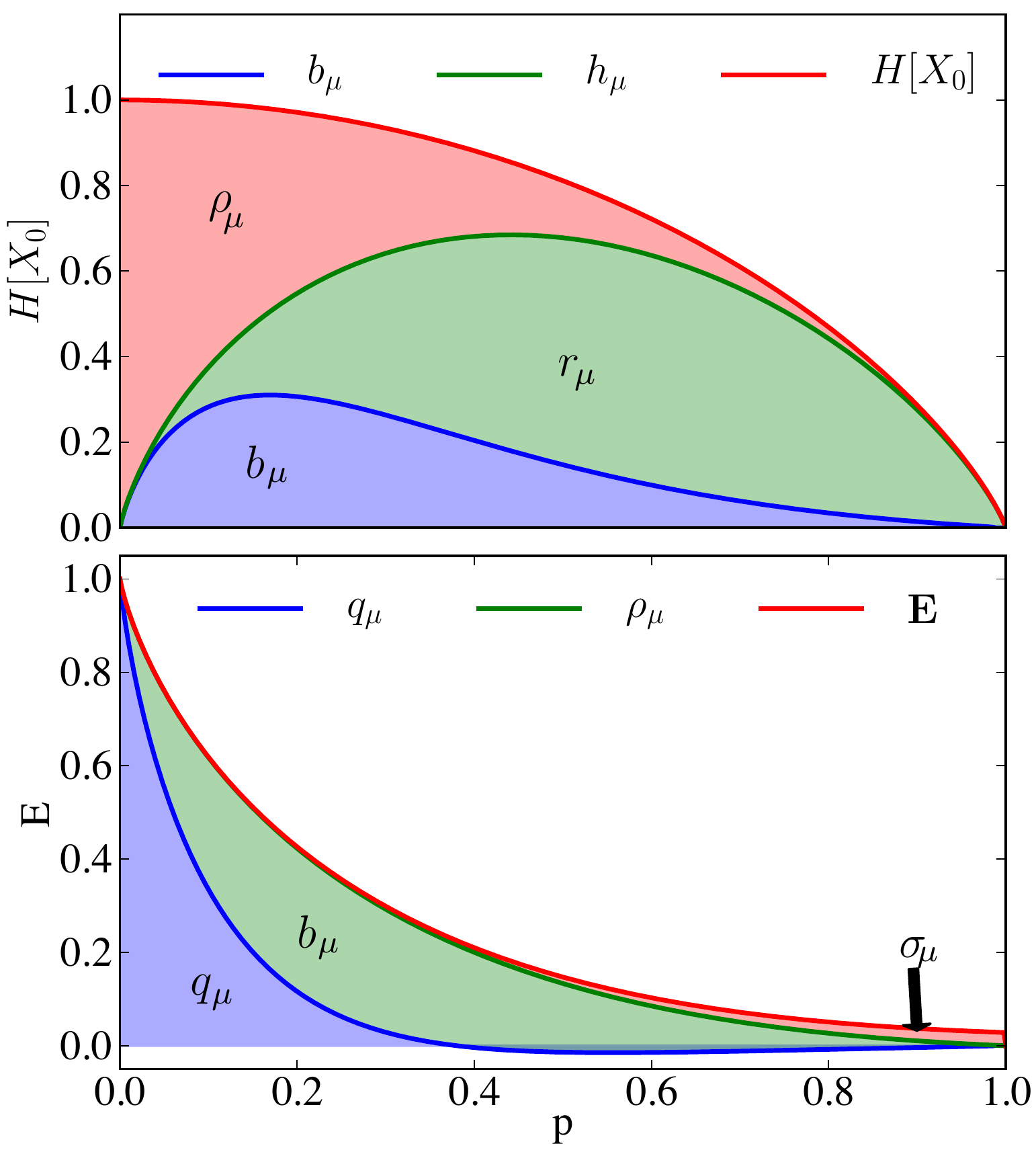}
\caption{(Top) Simple Nonunifilar Source information architecture as a function
  of $p$ with parameters $q=p$. The single-measurement entropy  $H[\MS_0]$ is
  the upper solid (red) line, entropy rate $\hmu$ the middle solid (green)
  line, the bound information $\bmu$ the lower solid (blue) line. Thus, the
  blue area corresponds to $\bmu$, the green area to the ephemeral information
  $\rmu=\hmu-\bmu$, and the red area to the single-symbol redundancy $\rhomu =
  H[\MS_0] - \hmu$. (Bottom) The components of the predictable
  information---the excess entropy $\EE=\sigmamu+\bmu+\qmu$ in bits---also as a
  function of $p$ with $q=p$. The lowest (blue) line is $\qmu$; the middle
  (green) line is $\qmu+\bmu$, so that the green area denotes $\bmu$'s
  contribution to $\EE$. The upper (red) line is $\EE$, so that the red area
  denotes elusive information $\sigmamu$ in $\EE$. Note that for a large range
  of $p$ the co-information $\qmu$ is (slightly) negative.
  }
\label{fig:SNS_InfoAnatomy}
\end{figure}

Using the formulae of Table \ref{tab:MeasuresFormulae}, Fig.
\ref{fig:SNS_Cmuetc} shows how the statistical complexity $\Cmu$, excess
entropy $\EE$, entropy rate $\hmu$, and bound information $\bmu$ vary with the
transition probabilities $p$ and $q$. $\Cmu$ often reveals detailed information
about a process' underlying structure, but for the parametrized SNS and other
renewal processes, the statistical complexity merely reflects the spread of the
interevent distribution. Thus, it increases with increasing $p$ and $q$. $\EE$,
a measure of how much can be predicted rather than historical memory required
for prediction, increases as $p$ and $q$ decrease. The intuition for this is
that as $p$ and $q$ vanish, the process arrives at a perfectly
predictable period-$2$ sequence. We see that the SNS constitutes a simple
example of a class of processes over which information transmission between the
past and future ($\EE$) and information storage ($\Cmu$) are anticorrelated.
The entropy rate $\hmu$ at the top right of Fig. \ref{fig:SNS_Cmuetc} is
maximized when transitions are uniformly stochastic and the bound information
$\bmu$ at the bottom right is maximized somewhere between fully stochastic and
fully deterministic regimes.

Figure \ref{fig:SNS_InfoAnatomy} presents a more nuanced decomposition of the
information measures as $p=q$ vary from $0$ to $1$. The top most plot breaks
down the single-measurement entropy $H[\MS_0]$ into redundant information
$\rhomu$ in a single measurement, predictively useless generated information $\rmu$,
and predictively useful generated entropy $\bmu$. As $p$ increases, the SNS
moves from mostly predictable (close to period-$2$) to mostly unpredictable,
shown by the relative height of the (green) line denoting $\hmu$ to the (red) line
denoting $H[\MS_0]$. The portion $\bmu$ of $\hmu$ predictive of the future is
maximized at lower $p$ when the single-measurement entropy is close to a less
noisy period-$2$ process. The plot at the bottom decomposes the predictable
information $\EE$ into the predictable information hidden from the present
$\sigmamu$, the predictable generated entropy in the present $\bmu$, and the
co-information $\qmu$ shared between past, present, and future.  Recall that
the co-information $\qmu = \EE-\sigmamu-\bmu$ can be negative and, for a large
range of values, it is. Most of the predictable information passes
through the present as indicated by $\sigmamu$ being a small for most
parameters $p$. Hence, even though the parametrized SNS is technically an
infinite-order Markov process, it can be well approximated by a
finite-order Markov process without much predictable information loss, as 
noted previously with rate-distortion theory \cite{Still10a}.


\section{Conclusions}
\label{sec:Conclusions}

Stationary renewal processes are well studied, easy to define, and, in many
ways, temporally simple. Given this simplicity and their long history it is
somewhat surprisingly that one is still able to discover new properties; in our
case, by viewing them through an information-theoretic lens. Indeed, their
simplicity becomes apparent in the informational and structural analyses. For
instance, renewal processes are causally reversible with isomorphic \eMs\ in
forward and reverse-time, i.e., temporally reversible. Applying the
causal-state equivalence relation to renewal processes, however, also revealed
several unanticipated subtleties.  For instance, we had to delineate new
subclasses of renewal process (``eventually $\Delta$-Poisson'') in order to
completely classify \eMs\ of renewal processes. Additionally, the informational
architecture formulae in Table \ref{tab:MeasuresFormulae} are surprisingly
complicated, since exactly calculating these informational measures requires a
unifilar presentation. In Sec. \ref{sec:SNS}, we needed an infinite-state
machine to study the informational architecture of a process generated by
simple two-state HMM.

Looking to the future, the new structural view of renewal processes will help
improve inference methods for infinite-state processes, as it tells us what to
expect in the ideal setting---what are the effective states, what are
appropriate null models, how informational quantities scale, and the like.
For example, Figs. \ref{fig:Generic_eM1}-\ref{fig:Generic_eM4} gave all
possible causal architectures for discrete-time stationary renewal processes.
Such a classification will allow for more efficient Bayesian inference of \eMs\
of point processes, as developed in Ref.  \cite{Stre13a}.  That is, we can
leverage ``expert'' knowledge that one is seeing a renewal process to delineate
the appropriate subset of model architectures and thereby avoid searching over
the superexponentially large set of all HMM topologies.

The range of the results' application is much larger than that explicitly
considered here. The formulae in Table \ref{tab:MeasuresFormulae} will be most
useful for understanding renewal processes with infinite statistical
complexity. For instance, Ref. \cite{Marz14e} applies the formulae to study the
divergence of the statistical complexity of continuous-time processes as the
observation time scale decreases. And, Ref. \cite{Marz14d} applies these
formulae to renewal processes with infinite excess entropy. In particular,
there we investigate the causal architectures of infinite-state processes that
generate so-called \emph{critical phenomena}---behavior with power-law temporal
or spatial correlations \cite{Crut93e}. The analysis of such critical systems
often turns on having an appropriate \emph{order parameter}. The statistical
complexity and excess entropy are application-agnostic order parameters
\cite{Crut97a,Feld02b,Tche13a} that allow one to better quantify when a phase
transition in stochastic processes has or has not occurred, as seen in Ref.
\cite{Marz14d}. Such critical behavior has even been implicated in early
studies of human communication \cite{Zipf35a} \footnote{Though see
\cite{Mand53a,Mill57a}.} and recently in neural dynamics \cite{Begg03a} and in
socially constructed, communal knowledge systems \cite{Dedeo13}.


\acknowledgments

The authors thank the Santa Fe Institute for its hospitality during visits and
C. Ellison, C. Hillar, R. James, and N. Travers for helpful comments. JPC is an
SFI External Faculty member. This material is based upon work supported by, or
in part by, the U.S. Army Research Laboratory and the U. S.  Army Research
Office under contracts W911NF-13-1-0390 and W911NF-12-1-0288. S.M. was funded
by a National Science Foundation Graduate Student Research Fellowship and the
U.C. Berkeley Chancellor's Fellowship.


\appendix

\section{Causal Architecture}
\label{app:CausalArchitecture}

{\Not Rather than write pasts and futures as semi-infinite sequences, we
notate a past as a list of nonnegative integers \cite{Cess13a}. The
semi-infinite past $\Past$ is equivalent to a list of interevent counts $N_{:0}$ and the count $N'_0$ since last event. Similarly, the semi-infinite
future $\Future$ is equivalent to the count to next event $N_0-N'_0$ and future
interevent counts $N_{1:}$.
} 

Now, recall Lemma \ref{lem:A}.

{\textbf{Lemma \ref{lem:A}}
\emph{The counts since last event are prescient statistics of a discrete-time stationary renewal process.}

{\ProLem This follows almost immediately from the definition of stationary
renewal process and the definition of causal states, since the random variables
$N_i$ are all i.i.d.. Then:
\begin{align*}
\Prob (\Future|\Past) = \Prob(N_0-N_0'|N_0') \prod_{i=1}^{\infty} \Prob(N_i)
  ~.
\end{align*}
And, therefore, $\Prob(\Future|\Past = \past) = \Prob(\Future|\Past=\past')$ is
equivalent to $\Prob(N_0-N_0'|N_0'=n_0) = \Prob(N_0-N_0'|N_0'=n_0')$. Hence, the counts since last event are prescient.
}

In light of Lemma \ref{lem:A}, we introduce new notation to efficiently refer
to groups of pasts with the same count since last event.

{\Not Let $r^+_n := \{\overleftarrow{x}: x_{-n-1:0} = 10^n\}$ for
$n\in\mathbb{Z}_{\geq 0}$. Recall that $10^n = 100 \cdots 00$, the sequence
with $n$ $0$s following a $1$.}

{\Rem Note that $\bm{\mathcal{R}}^+ = \{ r^+_n\}_{n=0}^{\infty}$ is always at least
a forward-time prescient rival, if not the forward-time causal states $\SSet^+$.
The probability distribution over $r^+_n$ is straightforward to derive. Saying
that $N_0'=n$ means there were $n$ $0$s since the last event, so that the
symbol at $X_{-n-1}$ must have been a $1$. That is:
\begin{align*}
\pi(r^+_n) & = \Prob(N_0'=n) \\
  & = \sum_{\ms\in\MeasAlphabet} \Prob(N_0'=n,X_{-n-1}=\ms) \\
  & = \Prob(N_0'=n,X_{-n-1}=1) \\
  & = \Prob(X_{-n-1}=1) \Prob(X_{-n:0}=0^n|X_{-n-1}=1)
  ~.
\end{align*}
Since this is a stationary process, $\Prob(X_{-n-1}=1)$ is independent of $n$,
implying:
\begin{align*}
\pi(r^+_n)  & \propto \Prob(X_{-n:0}=0^n|X_{-n-1}=1) \\
& = \sum_{m=0}^{\infty} \Prob(X_{-n:m+1}=0^{n+m}1|X_{-n-1}=1) \\
  & = \sum_{m=n}^{\infty} F(m) \\
  & = w(n)
  ~.
\end{align*}
We see that $\pi(r^+_n) = w(n) / Z$, with $Z$ a normalization constant that makes $\sum_{n=0}^{\infty} w(n) = \MIET$. And so:
\begin{align*}
\pi(r^+_n) = \frac{w(n)}{\MIET}
  ~.
\end{align*}
}

In the main text, Thm. \ref{thm:1} was stated with less precision so as to be
comprehensible. Here, we state it with more precision, even though the meaning
is obfuscated somewhat by doing so. In the proof, we still err somewhat on the
side of comprehensibility, and so one might view this proof as more of a proof
sketch.

\textbf{Theorem \ref{thm:1}}
\emph{The forward-time causal states of a discrete-time stationary renewal
process that is not eventually $\Delta$-Poisson are exactly $\SSet^+ =
\bm{\mathcal{R}}^+$, if $F$ has unbounded support.  When the support is bounded such
that $F(n) = 0$ for all $n\geq N$, $\SSet^+ = \{r^+_n\}_{n=0}^N$.  Finally, a discrete-time eventually $\Delta$-Poisson renewal process with characteristic
$(\tilde{n},\Delta)$ has forward-time causal states:
\begin{align*}
\SSet^+ = \{r^+_n\}_{n=0}^{\tilde{n}-1} \cup \{\cup_{k=0}^{\infty} r^+_{\tilde{n}+k\Delta+m}\}_{m=0}^{\Delta-1}
  ~.
\end{align*}
This is a complete classification of the causal states of any persistent renewal process.
}

{\ProThe
From the proof of Lemma \ref{lem:A} in this appendix, we know that two
prescient states $r^+_n$ and $r^+_{n'}$ are minimal only when:
\begin{equation}
\Prob(N_0-N_0'|N_0'=n) = \Prob(N_0-N_0'|N_0'=n')
  ~.
\end{equation}
Since $\Prob(N_0-N_0'=m|N_0'=n) = \Prob(N_0= m+n) / \Prob(N_0'=n)$,
$\Prob(N_0 = m+n) = F(m+n)$, and $\Prob(N_0'=n) = w(n) / (\MIET)$ from earlier,
we find that the equivalence class condition becomes:
\begin{equation}
\frac{F(m+n)}{w(n)} = \frac{F(m+n')}{w(n')}
  ~, \label{eq:eqClass}
\end{equation}
for all $m\geq 0$.

First, note that for these conditional probabilities even to be well defined,
$w(n) >0$ and $w(n')>0$. Hence, if $F$ has bounded support---$\max
\mathrm{supp}
F(n) = N$---then the causal states do not include any $r^+_n$ for $n>N$.
Furthermore, Eq.~(\ref{eq:eqClass}) cannot be true for all $m\geq 0$, unless
$n=n'$ for $n$ and $n'\leq N$. To see this, suppose that $n\neq n'$ but that
Eq.~(\ref{eq:eqClass}) holds. Then choose $m=N+1-\max (n,n')$ to give $0 =
F(N+1-|n-n'|) / w(n')$, a contradiction unless $n=n'$.

So, for all remaining cases, we can assume that $F$ in Eq.~(\ref{eq:eqClass}) has unbounded support.

A little rewriting makes the connection between Eq.~(\ref{eq:eqClass}) and
an eventually $\Delta$-Poisson process clearer. First, we choose $m=0$ to find:
\begin{align*}
\frac{F(n)}{w(n)} = \frac{F(n')}{w(n')},
\end{align*}
which we can use to rewrite Eq.~(\ref{eq:eqClass}) as:
\begin{align*}
\frac{F(m+n)}{F(n)} = \frac{F(m+n')}{F(n')},
\end{align*}
or more usefully:
\begin{equation*}
F(n'+m) = \frac{F(n')}{F(n)} F(n+m).
\end{equation*}
A particularly compact way of rewriting this is to define $\Delta' := n'-n$,
which gives $F(n'+m) = F((n+m)+\Delta')$.  In this form, it is clear that the
above equation is a recurrence relation on $F$ in steps of $\Delta'$, so that
we can write:
\begin{equation}
F((n+m)+k\Delta') = \left(\frac{F(n')}{F(n)}\right)^k F(n+m)
  ~.
\label{eq:eqClass2}
\end{equation}
This must be true for every $m\geq 0$. Importantly, since $w(n) =
\sum_{m=n}^{\infty} F(m)$, satisfying this recurrence relation is equivalent to
satisfying Eq.~(\ref{eq:eqClass}). But Eq.~(\ref{eq:eqClass2}) is just the
definition of an eventually $\Delta$-Poisson process in disguise; relabel with $\lambda :=
F(n') / F(n)$, $\tilde{n}:=n$, and $\Delta = \Delta'$.

Therefore, if Eq.~(\ref{eq:eqClass}) does not hold for any pair $n\neq n'$, the
process is not eventually $\Delta$-Poisson and the prescient states identified in Lemma
\ref{lem:A} are minimal; i.e., they are the causal states.

If Eq.~(\ref{eq:eqClass}) does hold for some $n\neq n'$, choose the minimal
such $n$ and $n'$ both.  The renewal process is eventually $\Delta$-Poisson
with characterization $\Delta = n'-n$ and $\tilde{n}$. And,
$F(\tilde{n}+m) / w(\tilde{n}+m) = F(\tilde{n}+m') / w(\tilde{n}+m')$ implies
that $m\equiv m'\mod\Delta$ since otherwise, the $n$ and $n'$ chosen would not
be minimal. Hence, the causal states are exactly those given in the theorem's statement.
}

{\Rem For the resulting $F(n)$ to be a valid interevent distribution, $\lambda
= F(\tilde{n}+\Delta) / F(\tilde{n}) < 1$ as normalization implies:
\begin{align*}
\sum_{n=0}^{\tilde{n}-1} F(n)
  + \sum_{n=\tilde{n}}^{\tilde{n}+\Delta-1}
  \frac{F(n)}{1-\lambda} = 1
  ~.
\end{align*}
}

{\Not Let's denote $\SSet^+ = \{\st^+_n := r^+_n\}_{n=0}^{\infty}$ for a renewal
process that is not eventually $\Delta$-Poisson,
$\SSet^+ = \{\st^+_n :=r^+_n\}_{n=0}^{\tilde{n}}$
for an eventually $\Delta$-Poisson renewal process with bounded support, and
$\SSet^+ = \{\st^+_n := r^+_n\}_{n=0}^{\tilde{n}-1} \cup
\{\st^+_{\tilde{n}+m} := \cup_{k=0}^{\infty}
r^+_{\tilde{n}+k\Delta+m}\}_{m=0}^{\Delta-1}$ for an
eventually $\Delta$-Poisson process.}

The probability distribution over these forward-time causal states is
straightforward to derive from $\pi(r^+_n) = w(n) / (\MIET)$. So, for a renewal
process that is not eventually $\Delta$-Poisson or one that is with bounded
support, $\pi(\st^+_n) = w(n) / (\MIET)$. (For the latter, $n$ only runs from
$0$ to $\tilde{n}$.) For an eventually $\Delta$-Poisson renewal process
$\pi(\st^+_n) = w(n) / (\MIET)$ when $n < \tilde{n}$ and:
\begin{align*}
\pi(\st^+_n) & = \sum_{k=0}^{\infty} \pi(r^+_{n+k\Delta}) \\
  & = \frac{\sum_{k=0}^{\infty} w(n+k\Delta)}{\MIET}
  ~,
\end{align*}
when $\tilde{n}\leq n <\tilde{n}+\Delta$.  And so, the statistical complexity given in Table \ref{tab:MeasuresFormulae} follows from $C_{\mu}^+ = H[\St^+]$.

Recall Lemma \ref{cor:ReversePrescientCounts} and Thm. \ref{cor:1}.

\textbf{Lemma \ref{cor:ReversePrescientCounts}}
\emph{Groupings of futures with the same counts to next event are reverse-time
prescient statistics for discrete-time stationary renewal processes.}

{\textbf{Theorem \ref{cor:1}}
\emph{(a) The reverse-time causal states of a discrete-time stationary renewal
process that is not eventually $\Delta$-Poisson are groupings of futures with the same count
to next event up until and including $N$, if $N$ is finite. (b) The reverse-time
causal states of a discrete-time eventually $\Delta$-Poisson stationary renewal process are groupings of futures with the same count to next event up until $\tilde{n}$, plus groupings of futures whose count since last event $n$ are in the same equivalence class as $\tilde{n}$ modulo $\Delta$.}
}

{\ProThe The proof for both claims relies on a single fact: In reverse-time, a
stationary renewal process is still a stationary renewal process with the same
interevent count distribution. The lemma and theorem therefore follow from
Lemma \ref{lem:A} and Thm. \ref{thm:1}.}

Since the forward and reverse-time causal states are the same with the same
future conditional probability distribution, we have $\FutureCmu = \PastCmu$ and the causal irreversibility vanishes: $\Xi = 0$.

Transition probabilities can be derived for both the renewal process's
prescient states and its \eM\ as follows. For the prescient machine, if a $0$
is observed when in $r^+_n$, we transition to $r^+_{n+1}$; else, we transition
to $r^+_0$ since we just saw an event. Basic calculations show that these transition probabilities are:
\begin{align*}
T^{(\ms)}_{r^+_n r^+_m}
  & = \Prob(\mathcal{R}^+_{t+1}=r^+_m,\MS_{t+1}=\ms|\mathcal{R}^+_t = r^+_n) \\
  & = \frac{F(n)}{w(n)}\delta_{m,0}\delta_{\ms,1}
    + \frac{w(n+1)}{w(n)}\delta_{m,n+1} \times \delta_{\ms,0}
  ~.
\end{align*}
Not only do these specify the prescient machine transition dynamic but, due to
the close correspondence between prescient and causal states, they
also automatically give the \eM\ transition dynamic:
\begin{align*}
T^{(\ms)}_{\st\st'}
  & = \Prob(\St^+_{t+1}=\st',\MS_{t+1}=\ms|\St^+_t=\st) \\
  & = \sum_{r,r' \in \bm{\mathcal{R}}^+ }
  T^{(\ms)}_{r'\rightarrow r} \Prob(\mathcal{S}^+_{t+1}=\st'|\mathcal{R}^+_{t+1}=r) \nonumber \\
  & \quad\quad \times \Prob(\mathcal{R}^+_{t}=r'|\mathcal{S}^+_{t} = \st)
  ~.
\end{align*}

\section{Information Architecture}
\label{app:InformationAnatomy}

It is straightforward to show that $\Prob(\MS_0 = 0 ) = \MIETI$ and, thus:
\begin{align*}
H[\MS_0] & = - \MIETI \log_2 \MIETI \\
  & \quad\quad - \Big( 1 - \MIETI \Big) \log_2 \Big( 1 - \MIETI \Big)
  ~.
\end{align*}
The entropy rate is readily calculated from the prescient machine:
\begin{align*}
\hmu & = \sum_{n=0}^{\infty}
  H[\MS_{t+1}|\mathcal{R}^+_t = r^+_n] \pi(r^+_n) \\
  & = - \sum_{n=0}^{\infty} \frac{w(n)}{\MIET}
  \Big( \frac{F(n)}{w(n)}\log_2 \frac{F(n)}{w(n)}  \\
  & \quad\quad + \frac{w(n+1)}{w(n)}\log_2 \frac{w(n+1)}{w(n)} \Big)
  ~.
\end{align*}
And, after some algebra, this simplifies to:
\begin{align*}
\hmu = - \MIETI \sum_{n=0}^{\infty} F(n)\log_2 F(n)
  ~,
\end{align*}
once we recognize that $w(0)=1$ and so $w(0)\log_2 w(0) = 0$ and we recall that
$w(n+1)+F(n)=w(n)$. The excess entropy, being the mutual information between
forward and reverse-time prescient states is \cite{Crut08a,Crut08b}:
\begin{align*}
\EE & = I[\mathcal{R}^+;\mathcal{R}^-] \\
    & = H[\mathcal{R}^+] - H[\mathcal{R}^+|\mathcal{R}^-]
  ~.
\end{align*}
And so, to calculate, we note that:
\begin{align*}
\Prob(r^+_n,r^-_m) & = \frac{F(m+n)}{\MIET} \text{~and~}\\
\Prob(r^+_n|r^-_m) & = \frac{F(n+m)}{w(m)}
  ~.
\end{align*}
After some algebra, we find that:
\begin{align*}
H[\mathcal{R}^+]  = - \sum_{n=0}^{\infty}
  \frac{w(n)}{\MIET} \log_2 \frac{w(n)}{\MIET}
\end{align*}
and that:
\begin{align*}
H[\mathcal{R}^+|\mathcal{R}^-]
  & = -\sum_{m,n=0}^{\infty}
  \frac{F(n+m)}{\MIET} \log_2 \frac{F(n+m)}{w(m)} \\
  & = -\sum_{m=0}^{\infty}
  \frac{m+1}{\MIET}F(m)\log_2 \frac{F(m)}{\MIET} \nonumber \\
  & \quad\quad + \sum_{m=0}^{\infty} \frac{w(m)}{\MIET}
  \log_2 \frac{w(m)}{\MIET}
  ~.
\end{align*}
The above quantity is the forward crypticity $\PC^+$ \cite{Crut08a} when the renewal process is not eventually $\Delta$-Poisson.
These together imply:
\begin{align*}
\EE & = -2\sum_{n=0}^{\infty}
  \frac{w(n)}{\MIET} \log_2 \frac{w(n)}{\MIET} \\
  & \quad + \sum_{m=0}^{\infty}
  (m+1) \frac{F(m)}{\MIET} \log_2 \frac{F(m)}{\MIET}
  ~.
\end{align*}

And, finally, the bound information $\bmu$ is:
\begin{align*}
\bmu & = I[\MS_{1:};\MS_0|\MS_{:0}] \\
     & = I[\mathcal{R}^-_{1};\MS_{0}|\mathcal{R}^+_0] \\
     & = H[\mathcal{R}^-_1|\mathcal{R}^+_0] - H[\mathcal{R}^-_1|\mathcal{R}^+_1]
  ~,
\end{align*}
where we used the causal shielding properties of prescient states,
$\MS_{:0}\rightarrow \mathcal{R}^+_0 \rightarrow \mathcal{R}^-_1 \rightarrow
\MS_{1:}$, and the unifilarity of the prescient machines as shown in Figs.
\ref{fig:Generic_eM1}-\ref{fig:Generic_eM4}.
While we already calculated $H[\mathcal{R}^-_1|\mathcal{R}^+_1]$, we
still need to calculate $H[\mathcal{R}^-_1|\mathcal{R}^+_0]$. We do so
using the prescient machine's transition dynamic. In particular:
\begin{align*}
\Prob( & \mathcal{R}^-_1=n | \mathcal{R}^+_0=m) \\
  & = \sum_{r \in \mathcal{R}^+}
  \Prob(\mathcal{R}^-_1=n|\mathcal{R}^+_1=r)
  \Prob(\mathcal{R}^+_1=r|\mathcal{R}^+_0=m) \\
  & = \frac{F(m+n+1)+F(n)F(m)}{w(m)}
  ~.
\end{align*}
Where we omit details getting to the last line.
Eventually, the calculation yields:
\begin{align*}
\bmu & = \frac{\sum_{n=0}^{\infty} (n+1)F(n)\log_2 F(n)}{\langle T\rangle} \\
     & \quad - \frac{\sum_{m,n=0}^{\infty} g(m,n)\log_2 g(m,n)}{\langle T\rangle}
  ~,
\end{align*}
where:
\begin{equation}
g(m,n) = F(m+n+1)+F(n)F(m)
  ~.
\end{equation}
From the expressions above, we immediately solve for $\rmu = \hmu - \bmu$,
$\qmu = H[\MS_0] - \hmu-\bmu$, and $\sigmamu = \EE - \qmu$. Thereby laying out
information architecture of stationary renewal processes.

Finally, we calculate the finite-time predictable information $\Ipred(M,N)$ as the mutual information between finite-time forward and reverse-time prescient states:
\begin{align}
\Ipred(M,N) = H[\mathcal{R}^{-_N}] - H[\mathcal{R}^{-_N}|\mathcal{R}^{+_M}]
  ~.
\label{eq:AppIpredMN}
\end{align}

Recall Corollary \ref{cor:2}.

{\textbf{Corollary \ref{cor:2}}
\emph{Forward-time (and reverse-time) finite-time $M$ prescient states of a discrete-time stationary renewal process are the counts from (and to) the next event up until and including $M$.}
}

{\ProCor
From Lemmas \ref{lem:A} and \ref{cor:ReversePrescientCounts}, we know that counts from (to)
the last (next) event are prescient forward-time (reverse-time) statistics. If
our window on pasts (futures) is $M$, then we cannot distinguish between
counts since (to) the last (next) event that are $M$ and larger. Hence, the
finite-time $M$ prescient statistics are the counts from (and to) the next
event up until and including $M$, where a finite-time $M$ prescient state
includes all pasts with $M$ or more counts from (to) the last (next) event.
}

To calculate $\Ipred(M,N)$, we find
$\Prob(\mathcal{R}^{+_M},\mathcal{R}^{-_N})$ by marginalizing $\Prob(\mathcal{R}^+,\mathcal{R}^-)$.
For ease of notation, we first define a function:
\begin{align*}
u(m) = \sum_{n=m}^{\infty} w(n)
  ~.
\end{align*}
Algebra not shown here yields:
\begin{align*}
\Ipred(M,N) & = H[\St^{-_N}]- H[\St^{-_N}|\St^{+_M}] \\
  & = \log_2 (\MIET) - \frac{\sum_{n=0}^{N-1} w(n)\log_2 w(n)}{\MIET} \\
  & \quad - \frac{\sum_{m=0}^{M-1} w(m)\log_2 w(m)}{\MIET} \\
  & \quad + \frac{\sum_{n=M}^{N+M-1} w(n)\log_2 w(n)}{\MIET} \\
  & \quad + \frac{\sum_{n=N}^{N+M-1} w(n)\log_2 w(n)}{\MIET} \\
  & \quad - \frac{u(N)\log_2 u(N) + u(M)\log_2 u(M)}{\MIET} \\
  & \quad + \frac{u(N+M)\log_2 u(N+M)}{\MIET} \\
  & \quad + \frac{\sum_{m=0}^{M-1} \sum_{n=m}^{N+m-1} F(n)\log_2 F(n)}{\MIET}
  ~.
\label{eq:AppIpredMNFinal}
\end{align*}
Two cases of interest are equal windows ($N=M$) and semi-infinite pasts
($M\rightarrow\infty$). In the former, we find:
\begin{align*}
\Ipred(M,M)
  & = \log_2 (\MIET) - \frac{2\sum_{m=0}^{M-1} w(m)\log_2 w(m)}{\MIET} \\
  & \quad + \frac{2\sum_{m=M}^{2M-1} w(m)\log_2 w(m)}{\MIET} \\
  & \quad - \frac{2u(M)\log_2 u(M)}{\MIET} + \frac{u(2M)\log_2 u(2M)}{\MIET} \\
  & \quad + \frac{\sum_{m=0}^{M-1} \sum_{n=m}^{M+m-1} F(n)\log_2 F(n)}{\MIET}
  ~.
\end{align*}
In the latter case of semi-infinite pasts several terms vanish and we have:
\begin{align*}
\Ipred(N)
  & = \log_2 (\MIET) - \frac{2 \sum_{n=0}^{N-1} w(n)\log_2 w(n)}{\MIET} \\
  & \quad - \frac{u(N)\log_2 u(N)}{\MIET} \\
  & \quad + \frac{N\sum_{n=N}^{\infty} F(n)\log_2 F(n)}{\MIET} \\
  & \quad + \frac{\sum_{n=0}^{N-1} (n+1) F(n)\log_2 F(n)}{\MIET}
  ~.
\end{align*}

\bibliography{chaos}

\end{document}